\documentclass[11pt,aps,showpacs,showkeys,preprintnumbers,amsmath,amssymb,nofootinbib]{revtex4}
\usepackage[toc,page]{appendix}
\usepackage{subfigure}
\usepackage{graphicx}
\usepackage{color}
\usepackage{amsmath}
\usepackage{amsfonts}
\usepackage{amssymb}
\usepackage{adjustbox}
\usepackage{multirow}
\def \be  {\begin{equation}}
\def \ee  {\end{equation}}
\def \ee  {\end{equation}}
\def \bea {\begin{eqnarray}}
\def \eea {\end{eqnarray}}

\begin{document}

\vspace*{1mm}

\title{Entropy per rapidity in Pb-Pb central collisions using Thermal and Artificial neural network(ANN) models at LHC energies}

\author{D. M. Habashy}
\affiliation{Ain Shams University, Faculty of Education, Physics Department, 11771, Roxy, Cairo, Egypt}

\author{Mahmoud Y. El-Bakry}
\affiliation{Ain Shams University, Faculty of Education, Physics Department, 11771, Roxy, Cairo, Egypt}

\author{Werner Scheinast}
\affiliation{Joint Institute for Nuclear Research - Veksler and Baldin Laboratory of High Energy Physics, Moscow Region, 141980 Dubna, Russia}

\author{Mahmoud Hanafy}
\email{mahmoud.nasar@fsc.bu.edu.eg}
\affiliation{Physics Department, Faculty of Science, Benha University, 13518, Benha, Egypt}

\date{\today}

\begin{abstract}

The entropy per rapidity $d S/d y$ produced in central Pb-Pb ultra-relativistic nuclear collisions at LHC energies is calculated using experimentally observed identified particle spectra and source radii estimated from Hanbury Brown-Twiss (HBT) for particles, $\pi$, $k$, $p$, $\Lambda$, $\Omega$, and $\bar{\Sigma}$, and $\pi$, $k$, $p$, $\Lambda$ and $K_s^0$ at $ \sqrt{s}$ $=2.76$ and $5.02$ TeV, respectively. Artificial neural network (ANN) simulation model is used to estimate the entropy per rapidity $d S/d y$ at the considered energies. The simulation results are compared with equivalent experimental data, and good agreement is achieved. A mathematical equation describes experimental data is obtained. Extrapolating the transverse momentum spectra at $p_T$ $=0$ is required to calculate $d S/d y$ thus we use two different fitting functions, Tsallis distribution and the Hadron Resonance Gas (HRG) model. The success of ANN model to describe the experimental measurements will imply further prediction for the entropy per rapidity in the absence of the experiment.  

\end{abstract}


\maketitle

\section{Introduction}
\label{sec:Intr}

Theoretical calculations by the lattice quantum chromodynamics (LQCD) approach  show that the quark-gluon plasma (QGP) phase, which is chirally restored and color deconfined, is formed at critical conditions of high energy density ($\epsilon \sim 1 $ GeV/$\mathtt{fm}^3$ ) and temperature ($ \mathtt{T} \sim 154 $ MeV) \cite{Karsch:2000kv,Pal:2003rz}.  
These conditions are expected to be obtained in Ultra-relativistic heavy ion collisions, when a dense medium of quarks and gluons is produced, which then experiences rapid-collective expansion  before the partons hadronize and subsequently decouple \cite{Pal:2003rz}. Many experiments are committed to discovering the QGP signals assuming quick thermalization, such as the Large Hadron Collider (LHC) at CERN, Geneva, and the Relativistic Heavy Ion Collider (RHIC) at BNL, USA \cite{Hanus:2019fnc, Busza:2018rrf}. Regrettably, measurements are limited to final state particles, the majority of which are hadrons \cite{Pal:2003rz}. The ensuing transverse and longitudinal expansion of the produced QGP is then studied by the relativistic viscous hydrodynamics models \cite{DerradideSouza:2015kpt}. In this case the net entropy, which is essentially conserved between preliminary thermalization until freeze-out \cite{Hanus:2019fnc, Busza:2018rrf,DerradideSouza:2015kpt,Pal:2003rz}, is an intriguing quantity which may provide significant information on the produced matter during the early stages of the nuclear collision. By accurately accounting for the entropy production at various phases of collisions, the observable particle multiplicities at the final state can be linked to system parameters, such as initial temperature, at earlier stages of the nuclear collisions \cite{Hanus:2019fnc}.

Two alternative methods are typically used to calculate the net created entropy during the collisions \cite{Hanus:2019fnc}. Pal and Pratt pioneered the first approach, which calculates entropy using transverse momentum spectra of various particle species and  their source sizes as calculated by HBT correlations \cite{Hanus:2019fnc,Pal:2003rz}. The original research analyzed experimental data taken from $\sqrt{s_{NN}} = 130$ GeV produced from Au-Au collisions and is still used to determine entropy at various energies \cite{Hanus:2019fnc, Busza:2018rrf}.
The second approach \cite{Sollfrank:1992ru,Muller:2005en} converts the multiplicity per rapidity $dN/dy$ produced at the final state to an entropy per rapidity $dS/dy$ using the entropy per hadron derived in a hadron resonance gas (HRG) model. Despite the fact that estimating the entropy per rapidity $ds/dy$ from the measured multiplicity $dN {ch} /d\eta$ is reasonably simple, the conversion factor between the measured charged-particle multiplicity dNch /d and the entropy per rapidity dS/dy in the literature \cite{Muller:2005en,Gubser:2008pc,Nonaka:2005vr,Berges:2017eom} is quite varied. Hanus and Reygers \cite{Hanus:2019fnc} estimated the entropy production using the transverse momentum distribution from data produced in p-p, and Pb-Pb collisions at $\sqrt{s} =$ $7$, and $2.76$ TeV for various particles , respectively.   
  
The present work aims to calculate the entropy per rapidity $ds/dy$ based on the transverse momentum distribution measured in Pb-Pb collisions at $\sqrt{s} =$ $2.76$, and $5.02$ TeV for particles $\pi$, $k$, $p$, $\Lambda$, $\Omega$, and $\bar{\Sigma}$, and $\pi$, $k$, $p$, $\Lambda$, and $K_s^0$, respectively. For a precise estimation of the entropy per rapidity $ds/dy$, we fitted the transverse momentum distribution of the considered particles using two thermal approaches, Tsalis distribution \cite{Cleymans:2016opp,Bhattacharyya:2017hdc} and the HRG model \cite{Yassin:2019xxl}. This enable us to cover a large range of the measured transverse particle momentum $p_{T}$, up to $\sim 20 $ GeV/c, unlike hanus that used a small range of $p_{T}$, $\sim 1.5 $ GeV/c and consider the particle's mass as free parameter. Indeed, we use the exact value of the particle's mass for all considered particle as in Particle Data Group (PDG) \cite{ParticleDataGroup:2018ovx}. Tsallis distribution succeeded to describe a large range of $p_{T}$ but cannot describe the whole range of $p_T$. That's why we use the HRG model to fit the other part of the $p_{T}$. Also, we estimated the entropy per rapidity $ds/dy$ for the considered particles using a very promising simulation model, the Artificial Neural Network (ANN). Recently, several modelling methods based on soft computing systems include the application of artificial intelligence (AI) techniques. These evolution algorithms have a physically powerful
existence in this field \cite{ar16,ar17,ar18,ar19,ar20}. The behavior of p-p and pb-pb interactions are complicated due to the non-linear relationship between the interaction parameters and the output. Understanding the interactions of fundamental particles requires multi-part data analysis and artificial intelligence techniques are vital. These techniques are useful as alternative approaches to conventional ones \cite{ar21}. In this sense, AI techniques such as Artificial Neural Network (ANN), Genetic Algorithm (GA), Genetic Programming (GP) and Genetic Expression Programming (GEP) can be used as alternative tools to simulate these interactions \cite{ar16, ar20}. The motivation for using an ANN approach is its learning algorithm, which learns the relationships between variables in data sets and then creates models to explain these relationships (mathematically dependant)\cite{ar29}. There is a desire for fresh computer science methods to analyze the experimental data for a better understanding of various physics phenomenons. ANNs have gained popularity in recent years as a powerful tool for establishing data correlations and have been successfully employed in materials science due to its generalisation, noise tolerance, and fault tolerance \cite{Annintro}. This enables us to use it to estimate the entropy per rapidity $ds/dy$. The results are then confronted to available experimental data and results obtained from previous calculation. 
 
The present paper is organised as follow. In Sec. (\ref{sec:models}), the used approaches are presented. Results and discussion are shown in Sec. (\ref{sec:Res}). The conclusion is drawn in Sec. (\ref{sec:Cncls}). A mathematical description for the entropy per rapidity $d s/d y$ and the transverse momentum spectra based on both Tsallis distribution and the HRG model are given in Appendices.

\section{The Used Approaches}
\label{sec:models}

In Sec. (\ref{sec:models}), we discuss the used methods for estimating the entropy per rapidity $ds/dy$ for various particles. The first method depends on the measured particle spectra for the considered particles \cite{Hanus:2019fnc}. In the second model, we use the ANN model, which may be considered the future simulation model \cite{Annintro}. 

\subsection{ Entropy per rapidity $ds/dy$ From Transverse Momentum distribution and HBT correlations}
\label{sec:spectra}

Here, we review the entropy per rapidity $d s/d y$ estimation from the phase space function distribution calculated from particle distribution spectra and femtoscopy \cite{Hanus:2019fnc}. The fandemetals of this approach are shown in Ref. \cite{Hanus:2019fnc,Bertsch:1994qc,Ferenc:1999ku}.

For any particles species at the thermal freeze-out stage, the entropy $S$ is obtained from the phase space distibution function $f (\vec{p} ,\vec{r})$ \cite{Hanus:2019fnc}

\begin{equation}
S=(2 J+1) \int \frac{d^{3} r d^{3} p}{(2 \pi)^{3}}[-f \ln f \pm(1 \pm f) \ln (1 \pm f)], \label{eq:1}
\end{equation}

where $+$ and $-$ stands for bosons and fermions, respectively. The quantity $2 J + 1$ represents the spin degeneracy for particles. The net entropy produced in the nuclear collisions is then obtained by summing over all the entropies of the created hadrons species. From Eq. (\ref{eq:1}), the integral can be expressed in a series expansion form 

\begin{equation}
\pm(1 \pm f) \ln (1 \pm f)=f \pm \frac{f^{2}}{2}-\frac{f^{3}}{6} \pm \frac{f^{4}}{12}+\ldots, \label{eq:2}
\end{equation}

The source radii, observed from HBT two particle correlations \cite{Lisa:2005dd} in three dimension, are calculated from  a longitudinally co-moving system (LCMS) where  the pair momentum component along the direction of the beam vanishes. In the LCMS, the source's density function is parametrized by a Gaussian in three dimension, allowing the phase space distribution function to be represented as \cite{Hanus:2019fnc}

\begin{equation}
f(\vec{p}, \vec{r})=\mathcal{F}(\vec{p}) \exp \left(-\frac{x_{\mathrm{out}}^{2}}{2 R_{\mathrm{out}}^{2}}-\frac{x_{\text {side }}^{2}}{2 R_{\text {side }}^{2}}-\frac{x_{\text {long }}^{2}}{2 R_{\text {long }}^{2}}\right), \label{eq:3}
\end{equation}

where $\mathcal{F}(\vec{p})$, the maximum phase density, is given by \cite{Hanus:2019fnc,Pal:2003rz}   

\begin{equation}
\mathcal{F}(\vec{p})=\frac{(2 \pi)^{3 / 2}}{2 J+1} \frac{d^{3} N}{d^{3} p} \frac{1}{R_{\text {out }} R_{\text {side }} R_{\text {long }}}, \label{eq:4}
\end{equation}

In Eqs. (\ref{eq:3}) and (\ref{eq:4}), the source radii are expressed in terms of the momentum $\vec{p}$.

Due to restricted statistics, in many circumstances only the source radius $R_{\mathrm{inv}}$ measured in one dimension, which is computed in the pair rest frame (PRF), may be obtained experimentally. 

The relationship between the PRF's $R_{\mathrm{inv}}$ and the three-dimensional source radii in the LCMS is considered to be Ref. \cite{Hanus:2019fnc,Pal:2003rz} 

\begin{equation}
R_{\mathrm{inv}}^{3} \approx \gamma R_{\mathrm{out}} R_{\mathrm{side}} R_{\mathrm{long}}, \label{eq:5}
\end{equation}

where $\gamma=m_{\mathrm{T}} / m \equiv \sqrt{m^{2}+p_{\mathrm{T}}^{2}} / m .$ 

 In Refs. \cite{Hanus:2019fnc,ALICE:2015tra} the ALICE collaboration published values for both $R_{\mathrm{inv}}$ and $R_{\mathrm{out}}, R_{\text {side }}, R_{\text {long }}$ determined from two pion correlations in Pb-Pb nuclear collisions at $\sqrt{s_{\mathrm{NN}}}=2.76 \mathrm{TeV}$.

 From these data Hanus et al., expressed a more general formula of Eq. (\ref{eq:5}) as \cite{Hanus:2019fnc}
 
\begin{equation}
R_{\text {inv }}^{3} \approx h(\gamma) R_{\text {out }} R_{\text {side }} R_{\text {long }}.   \label{eq:6} 
\end{equation}
 
 with $h(\gamma)=\alpha \gamma^{\beta}$. 

Form Eq. (\ref{eq:5}), the entropy per rapidity $d s/ d y$ can be given as \cite{Hanus:2019fnc}
\begin{equation}
\begin{aligned}
\frac{d S}{d y}=& \int d p_{T} 2 \pi p_{T} E \frac{d^{3} N}{d^{3} p}\left(\frac{5}{2}-\ln \mathcal{F} \pm \frac{\mathcal{F}}{2^{5 / 2}}\right.\\
&\left.-\frac{\mathcal{F}^{2}}{2 \times 3^{5 / 2}} \pm \frac{\mathcal{F}^{3}}{3 \times 4^{5 / 2}}\right), \label{eq:7}
\end{aligned}
\end{equation}
where $\mathcal{F}$, the phase space distribution function, is given by \cite{Hanus:2019fnc}
\begin{equation}
\mathcal{F}=\frac{1}{m} \frac{(2 \pi)^{3 / 2}}{2 J+1} \frac{1}{R_{\mathrm{inv}}^{3}\left(m_{\mathrm{T}}\right)} E \frac{d^{3} N}{d^{3} p}. \label{eq:8}
\end{equation}

For a better describtion for central $\mathrm{Pb}-\mathrm{Pb}$, Hanus et al., approximate expression $(1+f) \ln (1+f)$ in terms of Eq. (\ref{eq:1}) with numerical coefficients $a_{i}$ that is also used for high multiplicity values of $\mathcal{F}$ as \cite{Hanus:2019fnc}
\begin{equation}
\frac{d S}{d y}=\int d p_{T} 2 \pi p_{T} E \frac{d^{3} N}{d^{3} p}\left(\frac{5}{2}-\ln \mathcal{F}+\sum_{i=0}^{7} a_{i} \mathcal{F}^{i}\right). \label{eq:9}
\end{equation}

To calculate the entropy per rapidity $d s/ d y$ for the considered hadrons, the measured spectra of the transverse momentum $E \frac{d^{3} N}{d^{3} p}$ have to be extrapolated at $p_{T} = 0$. To do this, We confronted the $p_{T}$ momentum spectra to two various fitting functions estimated from two well-known models, Tsallis distribution and HRG model. A mathematical description of the transverse momentum distribution $E \frac{d^{3} N}{d^{3} p}$ using HRG model and Tsallis distribution is given in Appendices (\ref{sec:(append:hrg)}) and (\ref{sec:(append:tsalis)}), respectively.

\subsection{Artificial Neural Network(ANN) Model}
\label{sec:ann} 

\begin{figure}[htb]
\includegraphics[width=8cm]{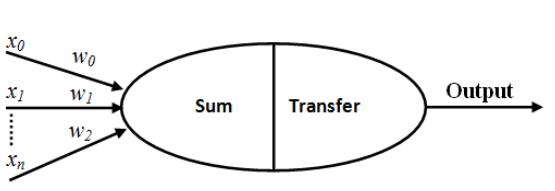}
\caption{Schematic diagram of a basic formal neuron.}
\label{fig:oneeai}
\end{figure}

Artificial neural network  model \cite{ar1,ar2,ar3,ar4,ar5,ar40,ar41} is a machine learning technique most popular in  high-energy physics community. In the last decade important physics results have been separated utilizing this model. Neuron is the essential processing component of Artificial neural network model (see Fig. \ref{fig:oneeai}), which forms a weighted sum of its input and passes the outcome to the yield through a non-linear transfer function. These transfer functions can also be linear, and then the weighted sum is sent directly to the output way. Eq. (\ref{eqa:1}) and Eq. (\ref{eqa:2}) represent respectively the weighted summation of the inputs and the non linear transfer function to the output of the neuron.

\begin{equation}
\sigma=\sum_{n} x_{n} w_{n}, \label{eqa:1}
\end{equation}

\begin{equation}
Y=f(\sigma), \label{eqa:2}
\end{equation}

\begin{figure}[htb]
\includegraphics[width=8cm]{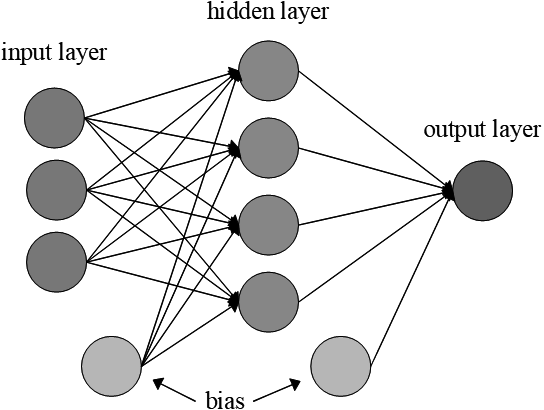}
\caption{Representative architecture of a feed-forward artificial neural network.}
\label{fig:twooai}
\end{figure}

The most widely recognized sort of ANN is multilayer feed forward neural network dependent on the BP (backpropagation) learning algorithm. Back propagation learning calculation is the most incredible in the Multi-layer calculation as shown in Alsmadi et al. \cite{ar6}. Multilayer feed-forward artificial neural network structure is a blend of various layers (see Fig. \ref{fig:twooai}). The primary layer (input layer) is the info layer that  presents the experimental  data then it is prepared and spread to the yield layer(output layer)through at least one hidden layer.

\begin{figure}[htb]
\includegraphics[width=8cm]{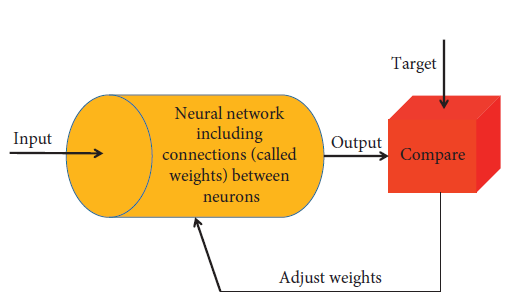}
\caption{Back-propagation network block diagram.}
\label{fig:threeeai}
\end{figure}

  Number of hidden layers and neurons required in every hidden layer are the important thing in designing a network. The best number of neuron and hidden layers rely upon many factors like the number of inputs, output of the network, the commotion of the target data, the intricacy of the error function, network design, and network training algorithm. In the greater part of cases, it is basically impossible to effortlessly decide the ideal number of hidden layers and number of neurons in each hidden layer without having to train the network. The training network comprises of constantly adjusting the weights of the association links between the processing as input patterns and required output components relating to the network. block diagram of the back propagation network is shown in Fig. \ref{fig:threeeai}. The aim of the training is to reduce and minimize the error which represents the difference between output experimental data $(t)$ and simulation results $(y)$ to accomplish the most ideal result.

Thus one tries to minimize the next mean square error (MSE) \cite{ar7}.

\begin{equation}
\mathrm{MSE}=\frac{1}{n} \sum_{i=1}^{n}\left(t_{i}-y_{i}\right)^{2}, \label{eqa:3}
\end{equation}

where n is  data points number used for training the model.

\subsubsection{Resilient propagation}

Resilient propagation, in short, RPROP\cite{ar8} is one of the quickest training algorithms available widely used for learning multilayer feed forward neural networks in numerous applications with the extraordinary advantage of basic applications. The RPROP algorithm simply alludes to the direction of the gradient. It is a supervised learning method. Resilient propagation calculates an individual delta $\Delta_{i j}$, for each connection, which determines the size of the weight update. The next learning rule is applied to calculate delta

\begin{equation} \label{eqa:4}
\begin{array}{c}
\Delta_{i j}{ }^{(t)}=\left\{\begin{array}{l}
\eta^{+} \times \Delta_{i j}{ }^{(t-1)} \quad, \text { if } \frac{\partial E}{\partial w_{i j}}^{(t-1)} \times \frac{\partial E^{(t)}}{\partial w_{i j}}>0 \\
\eta^{-} \times \Delta_{i j}{ }^{(t-1)}, \quad \text { if } \frac{\partial E}{\partial w_{i j}}^{(t-1)} \times \frac{\partial E^{(t)}}{\partial w_{i j}}<0 \\
\Delta_{i j}{ }^{(t-1)}, \text { else }
\end{array}\right. \\
\text { where } 0<\eta^{-}<1<\eta^{+}
\end{array}
\end{equation}

The update-amount $\Delta_{i j}$ develops during the learning process depend on the sign of the error gradient of the past iteration, $\frac{\partial E}{\partial w_{i j}}^{(t-1)}$ and the error gradient of the present iteration, $\frac{\partial E}{\partial w_{i j}}^{(t)}$. Each time the partial derivative (error gradient) of the corresponding weight $w_{i j}$ changes its sign, which shows that the last update too large and the calculation has jumped over a local minimum, the update-amount $\Delta_{i j}$ is decreased by the factor $\eta^{-}$ which is a constant usually with a value of $0.5$. If the derivative retains its sign, the update amount is slightly increased by the factor $\eta^{+}$ in order to accelerate convergence in shallow regions. $\eta^{+}$ is a constant usually with a value of 1.2. If the derivative is $0$, then we do not change the update-amount. When the update-amount is determined for each weight, the weight-update is then determined. 

The following equation is utilized to compute the weight-update

\begin{equation} \label{eqa:5}
\begin{array}{c}
\Delta w_{i j}^{(t)}=\left\{\begin{array}{l}
-\Delta_{i j}{ }^{(t)}, \text { if } \frac{\partial E^{(t)}}{\partial w_{i j}}>0 \\
+\Delta_{i j}{ }^{(t)},  \text { if } \frac{\partial E^{(t)}}{\partial w_{i j}}<0 \\
0,  \text { else } \\
w_{i j}^{(t+1)}  = w_{i j}{ }^{(t)}+\Delta w_{i j}{ }^{(t)}
\end{array}\right.
\end{array}
\end{equation}

If the present derivative is a positive amount meaning the past amount is also a positive amount (increasing error), then the weight is decreased by the update amount. If the present derivative is negative amount meaning the past amount is also a negative amount (decreasing error) then the weight is increased by the update amount.

\section{Results and Discussion}
\label{sec:Res}

In this section, we discussed the obtained results of the entropy per rapidity $d s/d y$ for central Pb-Pb at LHC energies, $\sqrt{s}$ $=2.76$ and $5.02$ TeV. The ANN simulation model is also used to estimate the entropy per rapidity $d s/d y$ at the considered energies. A comparison between the simulated results obtained from the experimental measurements and the simulated results is also shown.  
 
\subsection{The estimated Entropy per rapidity $ds/dy$ from Pb-Pb collisions at $\sqrt{s}$ $=2.76$ TeV }

We calculated the entropy per rapidity $d s/d y$, for particles $\pi$, $k$, $p$, $\Lambda$, $\Omega$ and $\bar{\Sigma}$, produced in central Pb-Pb collisions at $\sqrt{s}$ $=2.76$ TeV. The obtained results are compared to that estimated from the ANN simulation model and to that calculated in \cite{Hanus:2019fnc}. As experimental input, the computation includes transverse momentum spectra of the particles $\pi$, $k$, $p$ \cite{ALICE:2013mez}, $\Lambda$ \cite{ALICE:2013cdo}, $\Omega$ and $\bar{\Sigma}$ \cite{ALICE:2013xmt}. We also employ ALICE-measured HBT radii \cite{ALICE:2015hvw}. Also, Rprop based ANN is used to simulate $p_T$ spectra for the same particles. This procedure involves supervised learning algorithm that is implemented by using a set of input-output experimental data. As the nature of the output (various particles) is totally not the same, authors chose individual neural systems trained independently. Six networks are used to simulate different particles. Our networks have three inputs and one output. The inputs are $\sqrt{S}$, $P_{T}$ and Centrality. The output is {$\frac{1}{N_{evt} }\frac{d^2 N}{d y d p_T}$}.

\begin{figure}[htbp]
	\includegraphics[width=5cm]{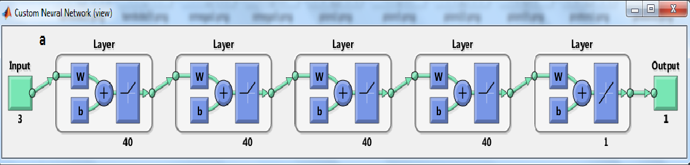}
	\includegraphics[width=5cm]{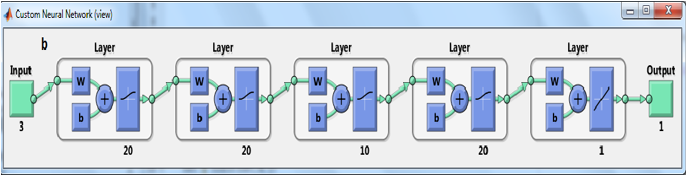}
	\includegraphics[width=5cm]{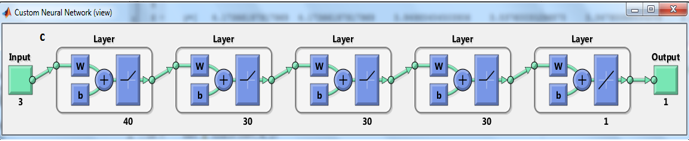}
	\includegraphics[width=5cm]{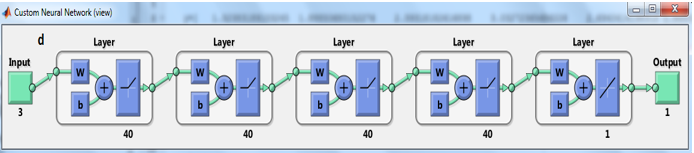}
	\includegraphics[width=5cm]{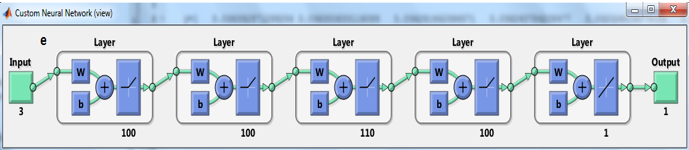}
	\includegraphics[width=5cm]{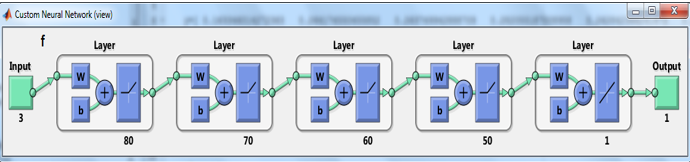}
	\caption{A schematic diagram of the basic formal neuron for particles (a) $\pi$, (b) $k$, (c) $p$, (d) $\Lambda$, (e) $\Omega$, and (f) $\bar{\Sigma}$.}
	\label{fig:oneeai1}
\end{figure}

Number of layers between input and output (hidden layer) and number of neurons in each hidden layer are selected by trial and error. In the beginning, we are begun with one hidden layer and one neuron in the hidden layer then the number of hidden layers and neurons are increased regularly. By changing the number of hidden layers and neurons, the performance of network would change. The learning performance of network can be measured and evaluated by inspecting the coefficients of the MSE and regression value (R). If the coefficient of the MSE is close to zero, it means that the difference between the network and desired output is small. Also, if it is zero, it means there is no difference or no error. On the other hand, R determines the correlation level between the output. And if it's value is equal to $1$, it means that the experimental results is compared with ANN model output and it has been found that there is a very good agreement between them. In our work, best MSE and R  values are obtained by using four hidden layers. The number of neurons in each hidden layer are ($40$, $40$, $40$, $40$), ($20$, $20$, $10$, $20$), ($40$, $30$, $30$, $30$), ($40$, $40$, $40$, $40$), ($100$, $100$, $110$, $100$), and ($80$, $70$, $60$, $50$) for particles $\pi$, $k$, $p$, $\Lambda$, $\Omega$, and $\bar{\Sigma}$, respectively. A simplification of the proposed ANN networks  are shown in Fig. \ref{fig:oneeai1} for particles  $\pi$ (a), $k$ (b), $p$ (c), $\Lambda$(d), $\Omega$(e), and $\bar{\Sigma}$(f) respectively.

\begin{figure}[htbp]
	\includegraphics[width=5cm]{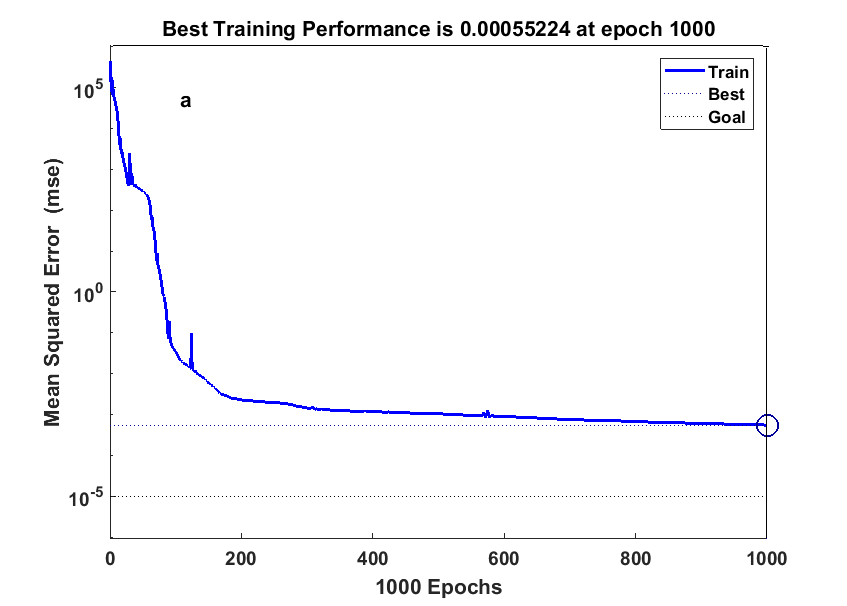}
	\includegraphics[width=5cm]{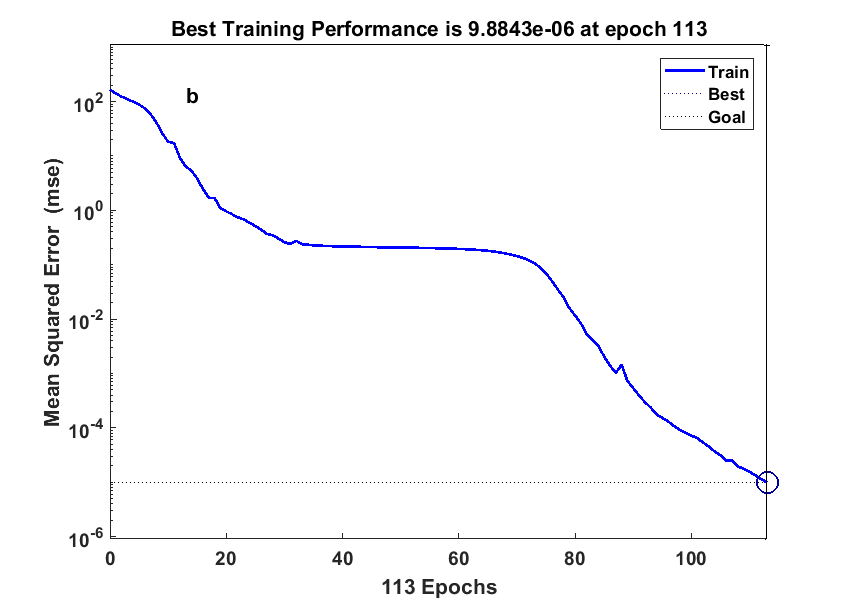}
	\includegraphics[width=5cm]{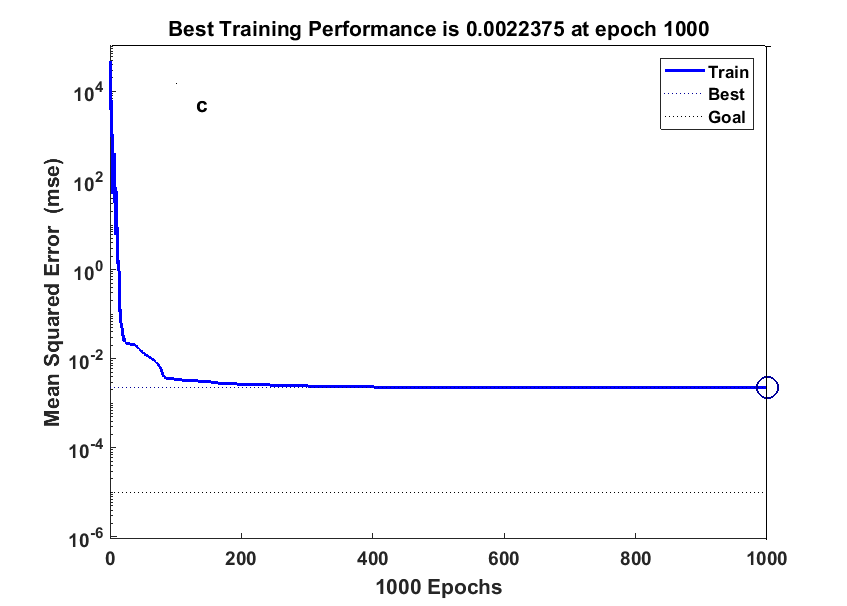}
	\includegraphics[width=5cm]{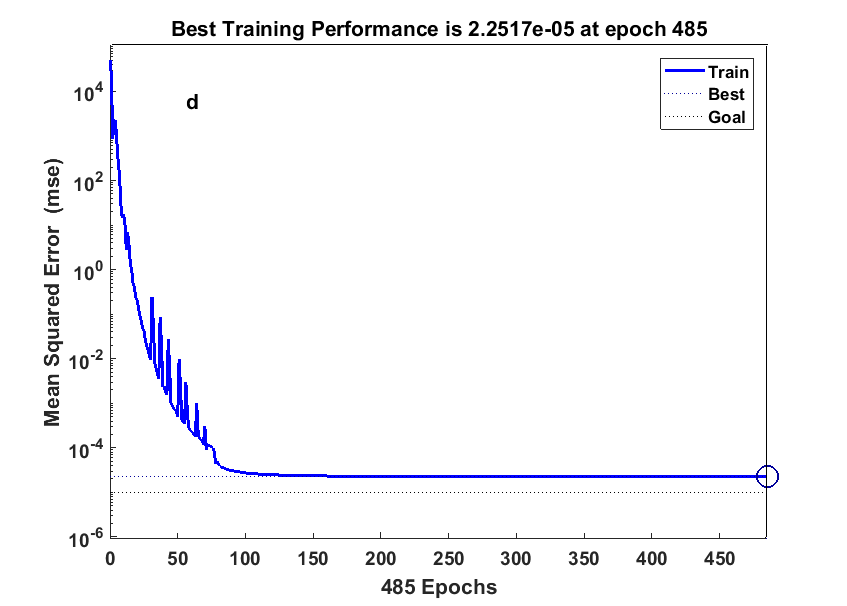}
	\includegraphics[width=5cm]{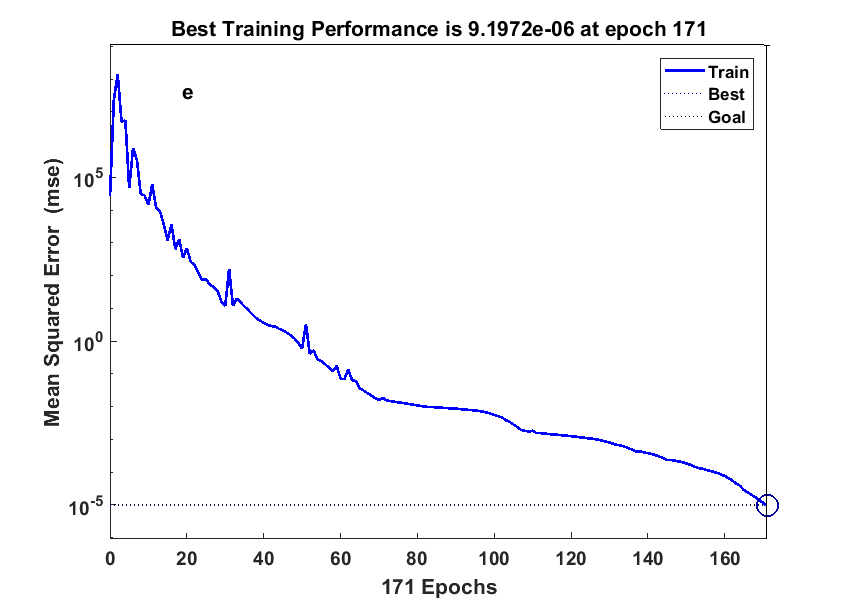}
	\includegraphics[width=5cm]{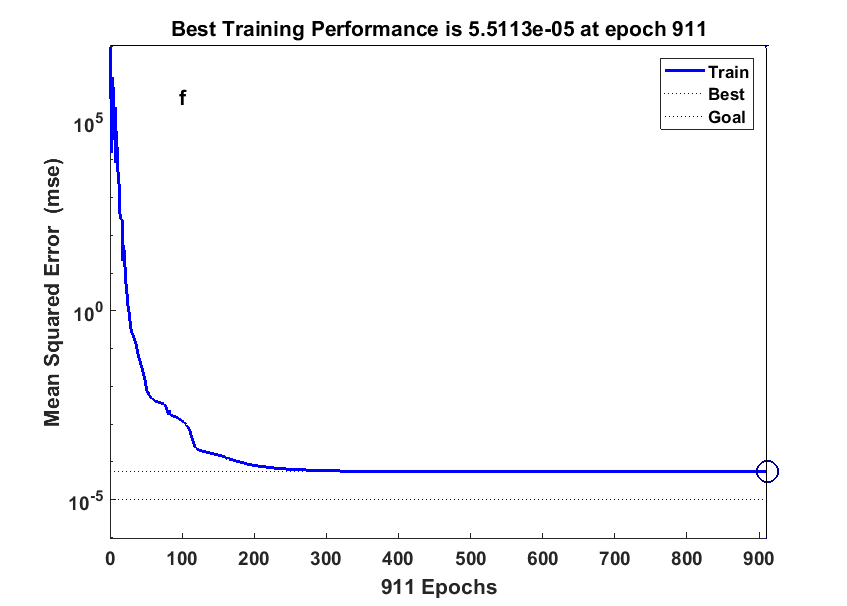}
	\caption{The best training performance (MSE) for particles (a) $\pi$, (b) $k$, (c) $p$, (d) $\Lambda$, (e) $\Omega$, and (f) $\bar{\Sigma}$.}
	\label{fig:oneeai2}
\end{figure}

\begin{figure}[htbp]
	\includegraphics[width=5cm]{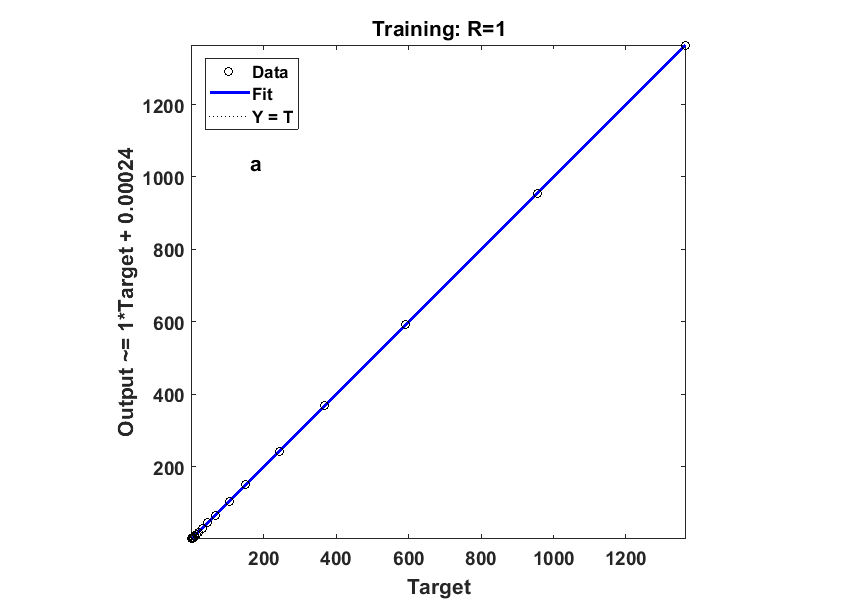}
	\includegraphics[width=5cm]{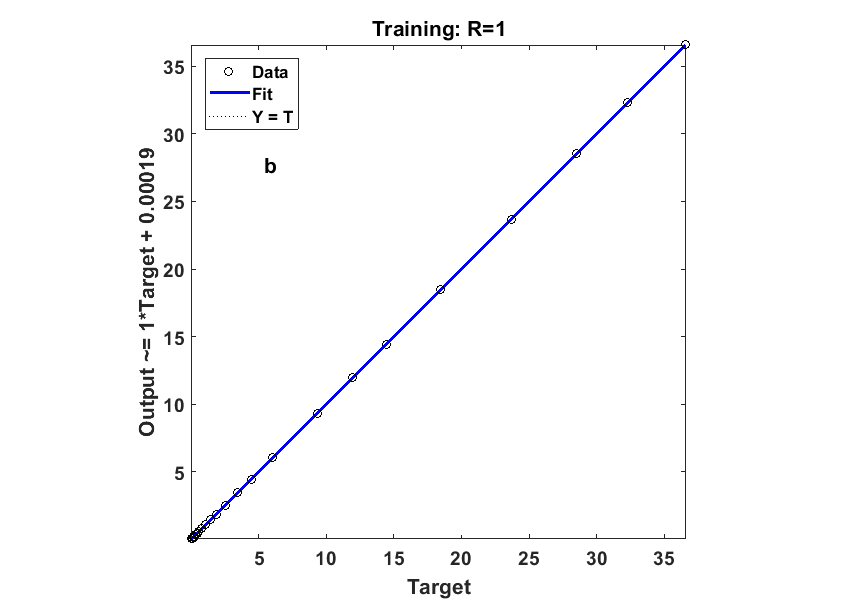}
	\includegraphics[width=5cm]{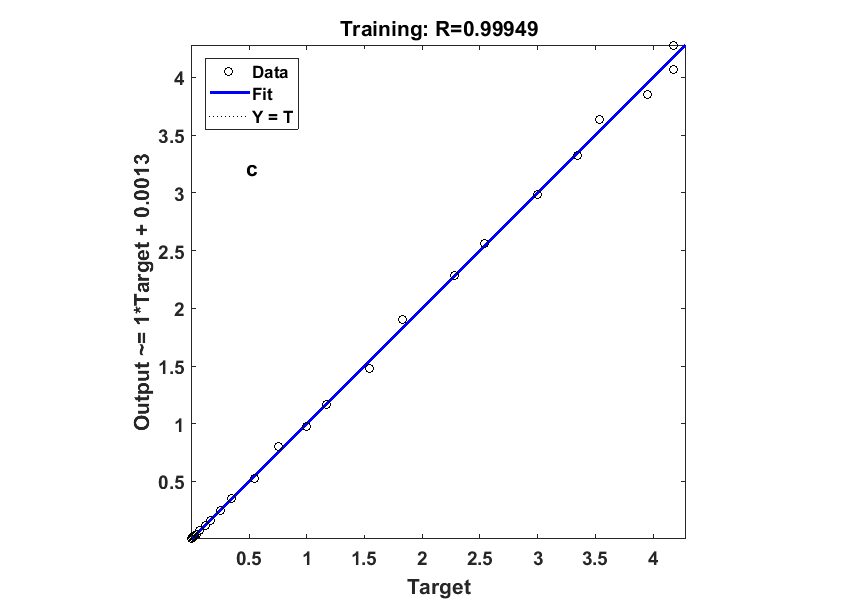}
	\includegraphics[width=5cm]{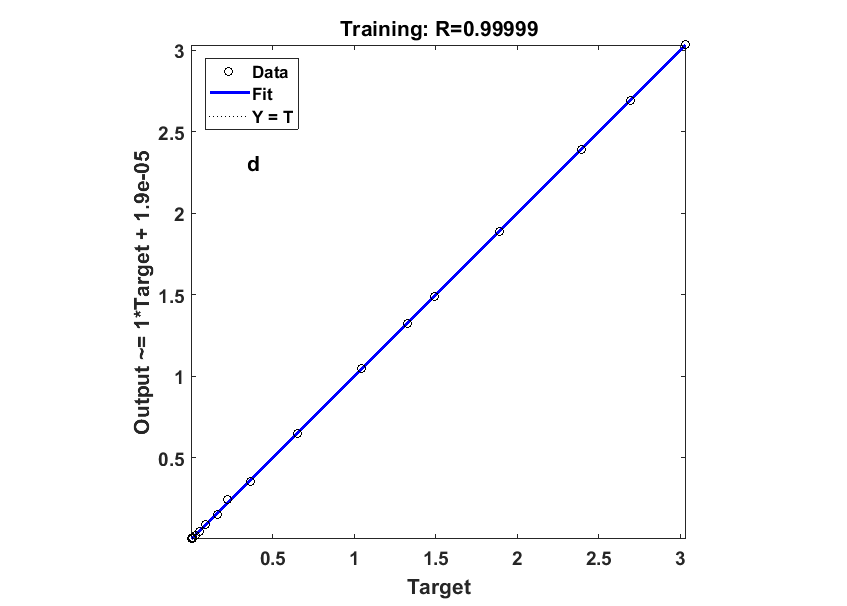}
	\includegraphics[width=5cm]{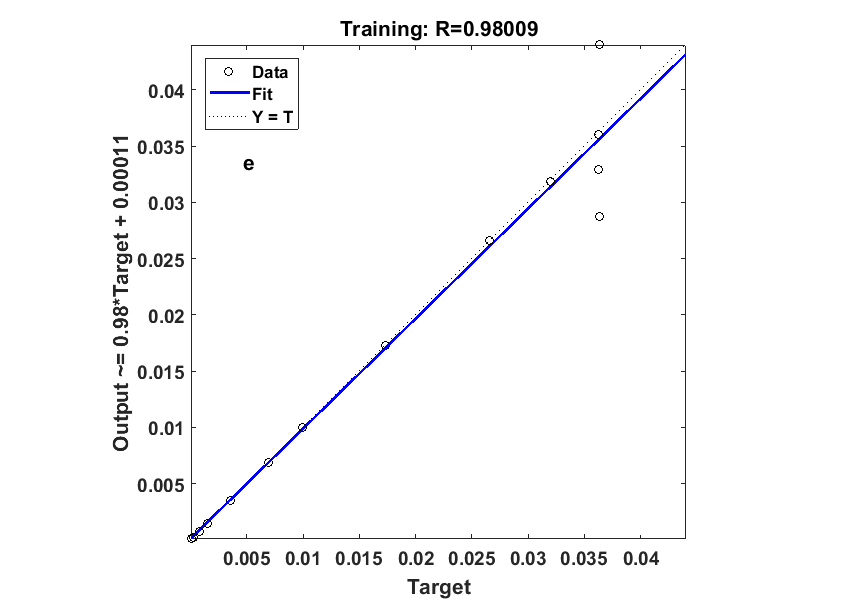}
	\includegraphics[width=5cm]{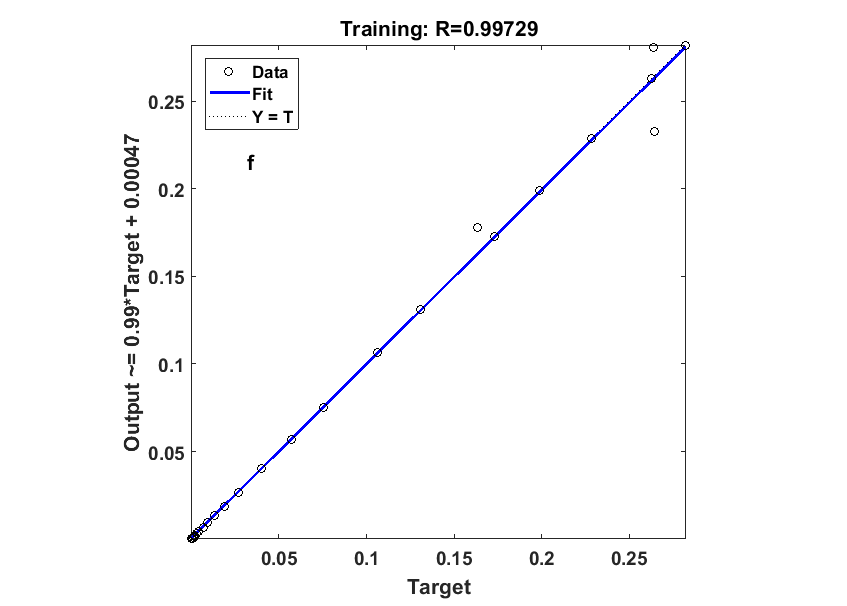}
	\caption{R Regression for particles (a) $\pi$, (b) $k$, (c) $p$, (d) $\Lambda$, (e) $\Omega$, and (f) $\bar{\Sigma}$ at the used epoch.}
\label{fig:oneeai3}
\end{figure}

The generated MSE and R for training  are shown in Figs. (\ref{fig:oneeai2}) and (\ref{fig:oneeai3}) for particles $\pi$ (a), $k$ (b), $p$ (c), $\Lambda$(d), $\Omega$(e), and $\bar{\Sigma}$(f), respectively. MSE values are $5.5224 \times 10^{-4}$, $9.8843 \times 10^{-6}$, $2.2375 \times 10^{-3}$, $2.2517 \times 10^{-5}$, $9.1972\times10^{-6}$ and $5.5113\times10^{-5}$  after epoch (number of training) $1000$, $113$, $1000$, $485$, $171$ and $911$ for particles $\pi$ (a), $k$ (b), $p$ (c), $\Lambda$(d), $\Omega$(e), and $\bar{\Sigma}$(f), respectively as in Fig. (\ref{fig:oneeai2}). Also, as shown in Fig.(\ref{fig:oneeai3}) regression values are closed to one. MSE and regression values mean good agreement between ANN results and experimental data. 

The transfer function used in hidden layer is \text{logsig} for $k$ particle and \text{poslin} for all other particles and \text{purelin} in output layer. All parameters used for ANN model are represented in Tab. (\ref{tabinputann276}). 

\begin {table}[htbp]
\caption {ANN parameters for particles $\pi$, $k$, $p$, $\Lambda$, $\Omega$, and $\bar{\Sigma}$ at $\sqrt{s}$ $=2.76$ TeV.}
\begin{adjustbox}{width=\columnwidth}
\begin{tabular}{|c|c|c|c|c|c|c|}
\hline
\multirow{2}{*}{ANN parameters} & \multicolumn{6}{|c|}{particles}\\
\cline { 2 - 7 }
 &$\pi$ & $K$ & $p$ & $\Lambda$ & $\Omega$ & $\bar{\Sigma}$ \\
\hline
Inputs & \multicolumn{2}{|c|}{$\sqrt{S} $}  &\multicolumn{2}{|c|}{$P_{T}(\mathrm{GeV})$}& \multicolumn{2}{|c|}{Centrality} \\
\hline
 $\sqrt{S}$ & \multicolumn{6}{|c|}{$2.76(\mathrm{TeV})$} \\
\hline
Output & \multicolumn{6}{|c|}{$\frac{1}{N_{evt} }\frac{d^2 N}{ d y d p_T}$} \\
\hline
Hidden layers & \multicolumn{6}{|c|}{4} \\
\hline
Neurons & $40,40,40,40$ & $20,20,10,20$ & $40,30,30,30$ & $40,40,40,40$ & $100,100,110,100$ & $80,70,60,50$ \\
\hline
Epochs & 1000 & 113 & 1000 & 485 & 171 & 911 \\
\hline
performance & $5.5224 \times 10^{-4}$  & $9.8843 \times 10^{-6}$ & $2.2375 \times 10^{-3}$& $2.2517 \times 10^{-5}$ & $9.1972 \times 10^{-6}$ & $5.5113 \times 10^{-5}$ \\
\hline
Training algorithms & \multicolumn{6}{|c|}{Rprop} \\
\hline
Training functions & \multicolumn{6}{|c|}{trainrp}\\
\hline
Transfer functions of hidden layers & Poslin & Logsig & Poslin & Poslin & Poslin & Poslin \\
\hline 
Output functions & \multicolumn{6}{|c|}{Purelin}\\
\hline
\end{tabular}
\end{adjustbox}
\label{tabinputann276}
\end {table}

To estimate the entropy $S$, extrapolation of the observed transverse momentum spectra to $p_T = 0$ is required. To achieve this, we fitted both the experimental and simulated $p_T$ spectra to two various functional models, Tsallis distribution \cite{Cleymans:2016opp,Bhattacharyya:2017hdc} and the HRG model \cite{Yassin:2019xxl}. The aim of using two different models is to fit the whole $p_T$ curve. 

\begin{figure}[htbp]
\includegraphics[width=8cm]{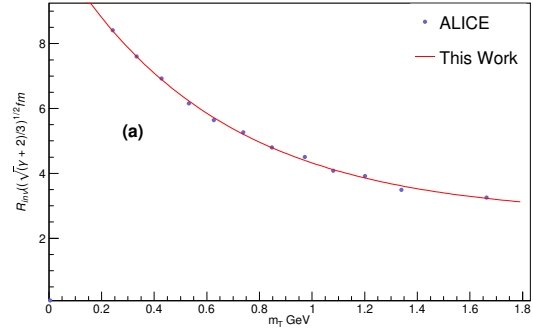}
\caption{The particle spectrum, measured by ALICE collaboration \cite{Kisiel:2014upa} and represented by closed blue circles symbols, is fitted to the Tasllis distribution \cite{Cleymans:2016opp,Bhattacharyya:2017hdc}, represented by solid red color, to extrapolate the spectrum at $p_{T} = 0$. The HBT one-dimensional radii scaled by $((2 + \gamma)/3)^{1/2}$ \cite{Hanus:2019fnc,Kisiel:2014upa} to be a function of transverse mass, $m_{T}$.}
\label{fig:oneer}
\end{figure}

\begin{figure}[htbp]
\includegraphics[width=5cm]{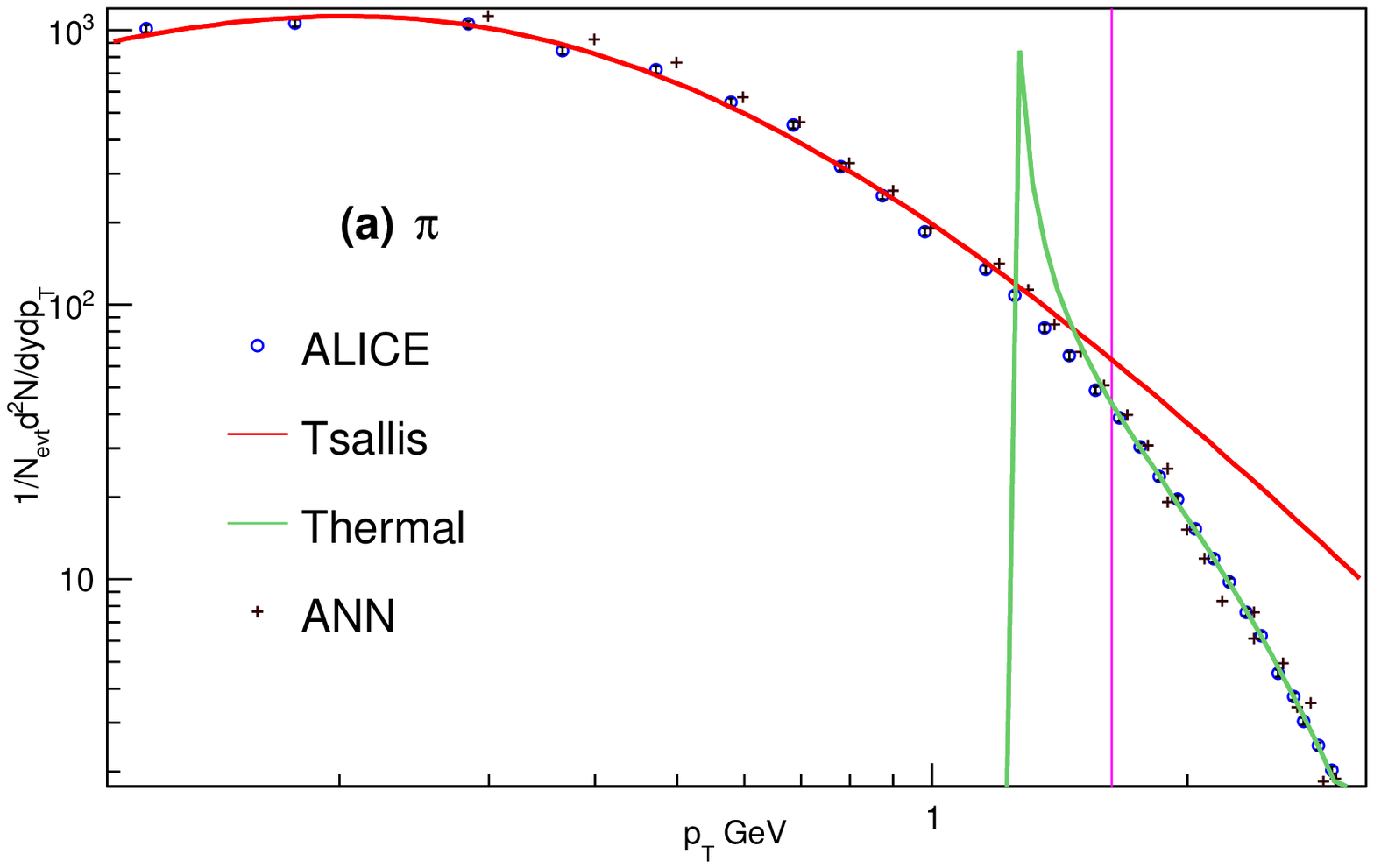}
\includegraphics[width=5cm]{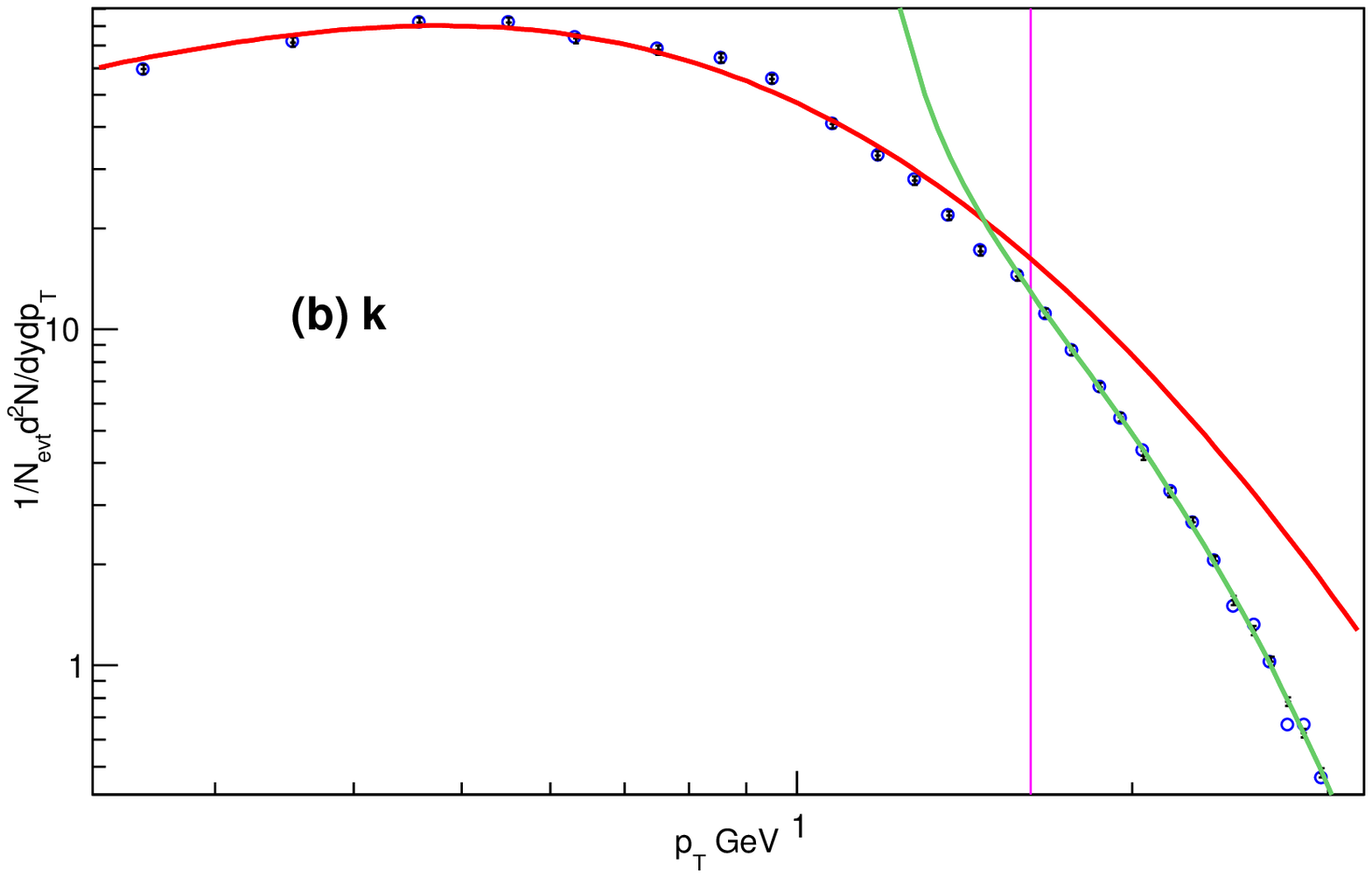}
\includegraphics[width=5cm]{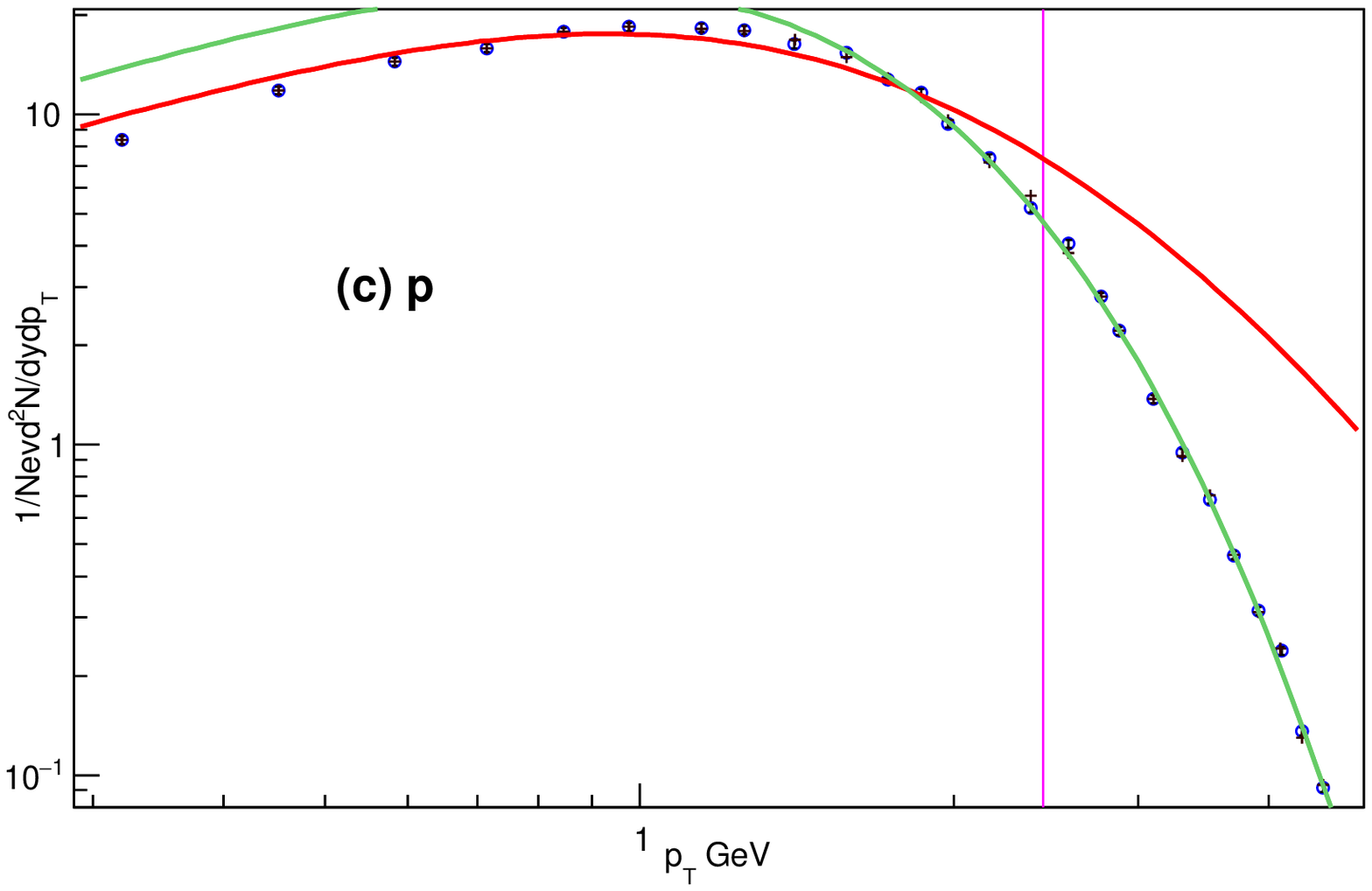}
\includegraphics[width=5cm]{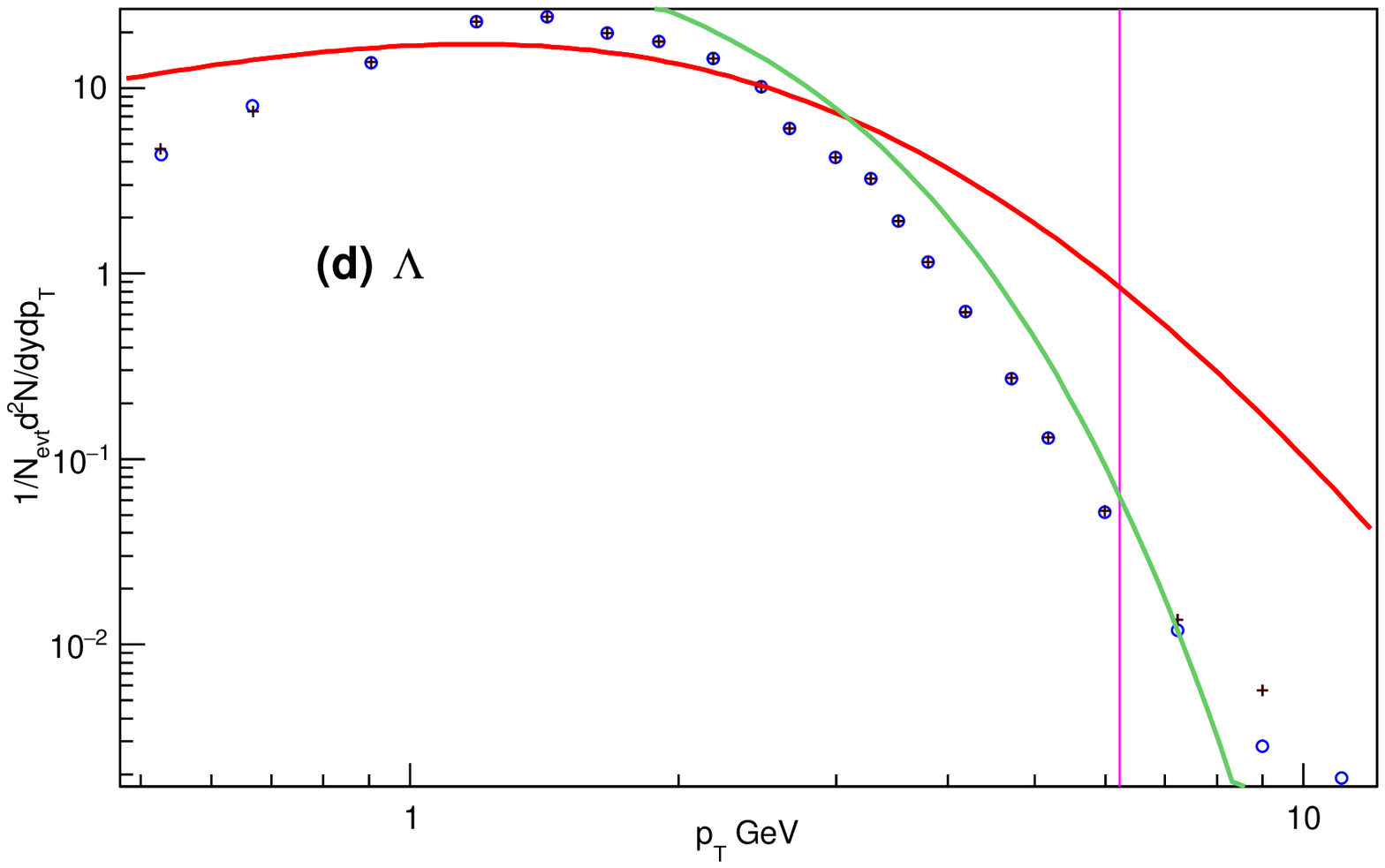}
\includegraphics[width=5cm]{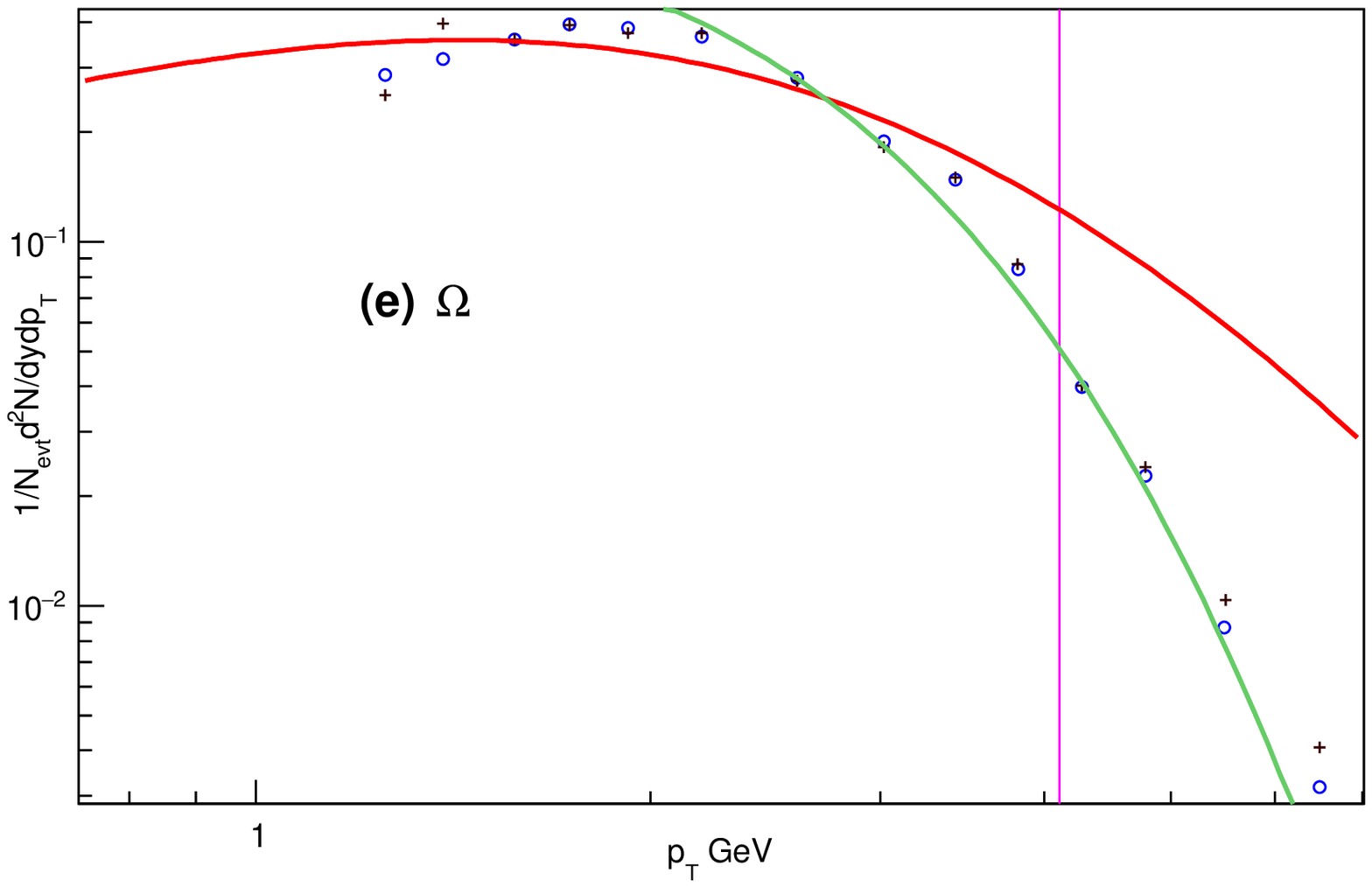}
\includegraphics[width=5cm]{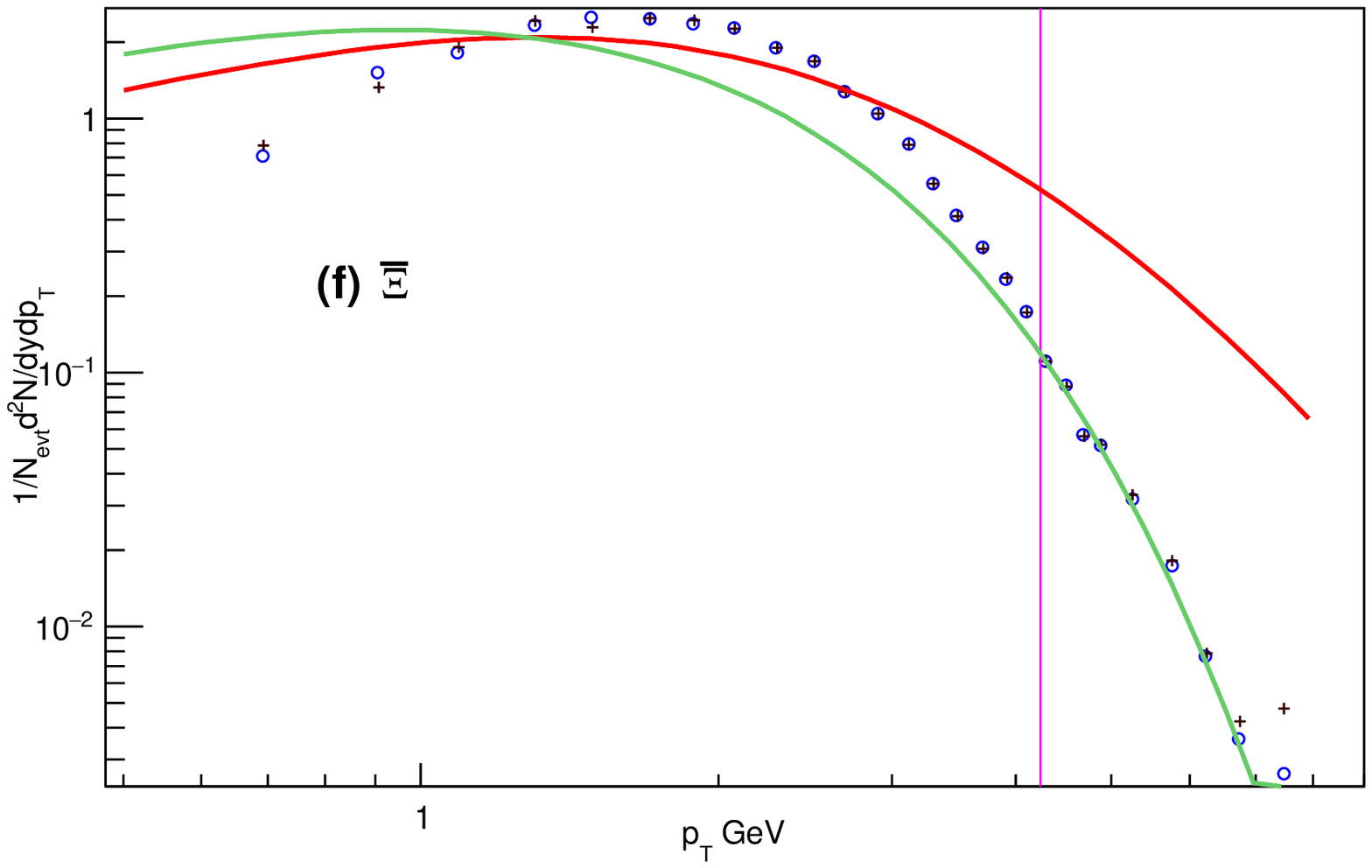}
\caption{The transverse momentum distribution, measured by ALICE experiment collaboration \cite{ALICE:2013mez,ALICE:2013cdo,ALICE:2013xmt} at centre of mass energy $= 2.76$ TeV and represented by blue open circles symbols, for particles $\pi$ (a), $k$ (b), $p$ (c), $\Lambda$ (d), $\Omega$ (e), and $\bar{\Sigma}$ (f) is compared to the statistical fits from Tsallis distribution, perfectly fits at $0.001 < y < 6 $, represented by red solid line given by Eq. (\ref{eqb:5}) and the HRG model, works in $ 6 < y < 12 $, represented by solid green line given by Eq. (\ref{eqq(7)}). A boarder line is drawn between Tsallis and HRG models and represented by solid purple color. The experimental data and the results of both models are then confronted to that obtained from the ANN simulation model, represented by dark brown plus sign symbols.}
\label{fig:twon}
\end{figure}

Fig. (\ref{fig:oneer}) shows the particle spectrum, measured by ALICE collaboration \cite{Kisiel:2014upa} and represented by closed blue circles symbols, is fitted to the Tasllis distribution \cite{Cleymans:2016opp,Bhattacharyya:2017hdc}, represented by solid red color, to extrapolate the spectrum at $p_{T} = 0$. The HBT one-dimensional radii scaled by $((2 + \gamma)/3)^{1/2}$ \cite{Hanus:2019fnc,Kisiel:2014upa} to be a function of transverse mass, $m_{T}$. Confronting both the experimental and simulated particle spectra $p_T$ to both Tsallis distribution and HRG model are shown in Fig. (\ref{fig:twon}) for particles $\pi$ (a), $k$ (b), $p$ (c), $\Lambda$ (d), $\Omega$ (e), and $\bar{\Sigma}$ (f). It is clear from Fig. (\ref{fig:twon}) that using the various forms of the fitting function is obvious as the Tsallis function can fit only the left side of the $p_{T}$ curve at $0.001 < y < 6 $ while the HRG model can fit the right side $ 6 < y < 12 $ as well. This conclusion can encourage us for further investigation. The obtained fitting parameters as a result of both Tsallis distribution and HRG model are summarized in Tabs. (\ref{tab1exfitt276}) and (\ref{tab1annfitt276}), respectively.

\begin {table}[htbp]
\caption {The transverse momentum distribution fitting parameters when confronting the Tasllis distribution, Eq. (\ref{eqb:5}), and the HRG model, Eq. (\ref{eqq(7)}), to the ALICE experimental data \cite{ALICE:2013mez,ALICE:2013cdo,ALICE:2013xmt} at $\sqrt{s}$ $= 2.76$ TeV for particles $\pi$, $k$, $p$, $\Lambda$, $\Omega$, and $\bar{\Sigma}$.}
\begin{adjustbox}{width=\columnwidth}
\begin{tabular}{|c|c|c|c|c|c|c|c|}
\hline 
 \multirow{2}{*}{particle} & \multicolumn{3}{c|}{Tsallis distribution}&\multicolumn{3}{c|}{HRG model} & \multirow{2}{*}{$\chi^{2}$ /dof} \\ 
 \cline{2-7}
  &  dN/dy  & $T_{Ts}$ GeV & q &V $fm^3$  &$T_{th}$ GeV & $\mu$ GeV &  \\
\hline 
 $\pi$&  $739.886$ & $0.0658$ & $1.2305$ & $3.41332 \times 10^1 \pm  6.103$ &$3.3881\times 10^{-1} \pm 4.8231 \times 10^{-3}$ & $ 1.2549 \pm 4.219 \times 10^{-2}$&  $13.6/12$  \\ 
\hline 
   $K$ & $88.3303$  & $0.1711$ &$1.1132$  &$1.38260 \times 10^1 \pm  3.6055 $ &$3.13028\times 10^{-1} \pm  4.36637\times 10^{-3}$ & $1.26516 \pm  6.32063\times 10^{-2}$&     $163.536/11$ \\ 
\hline 
  $p$ & $39.8814$  &$0.31752$  &$1.13739$  &  $5.57340\times 10^2 \pm  3.31936\times 10^{2}$& $3.89646\times 10^{-1}\pm   1.79383\times 10^{-3} $& $3.10152\times 10^{-2} \pm  2.39531\times 10^{-1}$ &  $ 25.3567/12$   \\ 
\hline
   $\Lambda$ & $47.6767$  & $0.4388$ & $1.1148$ &$48.7847   \pm   48983.2$ &  $0.503582   \pm   45.8599$& $1.12747 \pm   626.183 $&   $286.085/15$   \\ 
\hline
   $\Omega $& $1.24793$  & $0.4277$ & $1.14736$ &$1.24168   \pm   1.0661$ &$ 0.539524 \pm   0.0513$ & $1.14468  \pm   0.3706$&   $0.0199/7$   \\ 
\hline  
 $\bar{\Sigma}$& $6.5279$  & $0.4640$ & $1.1245$ & $4.38896 \pm   1.5158$ &$0.545919  \pm  0.0209$ & 
 $0.865  \pm   0.1993$& $ 3.0475/15 $      \\ 
\hline  
\end{tabular} 
\end{adjustbox}
\label{tab1exfitt276}
\end {table}

\begin {table}[htbp]
\caption {The same in Tab. (\ref{tab1exfitt276}) but the statistical fits results, from both used models, is confronted to the ANN simulation model.}
\begin{adjustbox}{width=\columnwidth}
\begin{tabular}{|c|c|c|c|c|c|c|c|}
\hline 
 \multirow{2}{*}{particle} & \multicolumn{3}{c|}{Tsallis distribution}&\multicolumn{3}{c|}{HRG model} & \multirow{2}{*}{$\chi^{2}$ /dof} \\ 
 \cline{2-7}
  & $ dN/dy$  & $T$ GeV & q &V $fm^3$  &$T$ GeV & $\mu$ GeV &  \\
\hline 
 $\pi$& $775.496$  & $0.08174$ & $1.1873$ & $23.5314 \pm 15.5048$  & $0.3286 \pm 0.0448$&$1.3797 \pm 0.09944$   
 & $32.0297/11 $ \\ 
\hline 
   $K$ &$88.2811$ &$0.1719$  &$1.111$  &$13.8111 \pm 1.993$  & $ 0.3125 \pm 0.00478$ &$ 1.2664 \pm 0.0292$ &$  159.234/11 $  \\ 
\hline 
  $p$ &$40.2232$ & $0.31542$ & $1.14088$ &$26.0248 \pm 3.2175$ & $0.3758 \pm 0.0053$ &$ 1.2924 \pm 0.0486$& $  23.8924/12 $ \\ 
\hline
   $\Lambda$ & $47.953$ & $0.4458$ &$1.11257$  &$ 50.852 \pm 54211.8$&$0.5057 \pm 30.6755 $&$1.1584 \pm 358.797 $ &$   285.464/15 $  \\ 
\hline
   $\Omega $& $1.2580$ & $0.4125$ & $1.1492$ & $1.17248 \pm 0.294$&$0.6464 \pm 0.0184$&$0.4864 \pm 0.1411$ &$0.02487/7 $ \\ 
\hline  
 $\bar{\Sigma}$& $6.5309$ & $0.4595$ & $1.1260$ &$4.3795 \pm 1.8017$& $0.5577 \pm 0.0248$ &$0.7849 \pm 0.2178$ &$ 3.1109/15$ \\ 
\hline  
\end{tabular} 
\end{adjustbox}
\label{tab1annfitt276}
\end {table}

The estimated entropy per rapidity $ds/dy$ from Pb-Pb central collisions at $\sqrt{s}$ $=2.76$ TeV using the Tsallis distribution, HRG model, and the ANN model for particles $\pi$, $k$, $p$, $\Lambda$, $\Omega$ and $\bar{\Sigma}$ is represented in Tab. (\ref{tab3276entropy}). The effect of both the Tsallis distribution and HRG model fitting function on the estimated entropy per rapidity $d s/d y$ is also shown in Tab. (\ref{tab3276entropy}). We compare the entropy per rapidity obtained from the statistical fits and ANN model to that obtained in Ref. \cite{Hanus:2019fnc}. The function which describes the non-linear relationship between inputs and output based ANN simulation model is given in Appendix \ref{sec:(append:neural)}. The results of ANN simulation, Tsallis distribution and the HRG model for particles compared with experimental data are shown in Fig.(\ref{fig:twon}).

\begin {table}[htbp]
\caption {The estimated entropy per rapidity $ds/dy$ from Pb-Pb central collisions at $\sqrt{s}$ $=2.76$ TeV using the Tsallis distribution, HRG model, and the ANN model. The obtained results are compared to that obtained in Ref. \cite{Hanus:2019fnc}.}
\begin{adjustbox}{width=\columnwidth}
\begin{tabular}{|c|c|c|c|c|c|}
\hline 
 particle & $(ds/dy)_{y=0}$ & $(ds/dy)_{y=0}$ supplemented by Tsallis & $(ds/dy)_{y=0}$ supplemented by HRG model & $(ds/dy)_{y=0}$ estimated by ANN model  & $(ds/dy)_{y=0}$ Ref. \cite{Hanus:2019fnc}  \\ 
\hline 
$\pi$ & $1908.21$  & $2260.85$ & $ 2267.58$ &$2265.17$ &  $ 2182$  \\ 
\hline 
  $K$ & $478.351 $ & $512.399 $& $514.321$ &$514.347 $& $605 $  \\ 
\hline 
$p$ & $265.648$  & $ 278.125$& $278.486 $& $277.937$& $   266$   \\ 
\hline
  $\Lambda$ & $304.334 $ &$325.742 $ & $321.939$ & $300 $& $ 320  $ \\ 
\hline
  $\Omega $ &$10.1561$   & $14.4025$ & $14.2159$ &$13.3129$ & $ 16$    \\ 
\hline  
 $\bar{\Sigma}$& $54.3717 $ & $58.3102 $& $57.8227 $&$58.1449$ &  $58 $  \\ 
\hline  
\end{tabular} 
\end{adjustbox}
\label{tab3276entropy}
\end {table}

From Tab. (\ref{tab3276entropy}), The calculated entropy per rapidity $d s/d y$ form the statistical fits, ANN model and that obtained in Ref. \cite{Hanus:2019fnc} are agree with each other. The excellent agreement between the estimated results of $d s/d y$ from ANN simulation model and to that obtained in Ref. \cite{Hanus:2019fnc} encourage us to use it at another energies.

\subsection{The estimated entropy per rapidity from Pb-Pb collisions at $\sqrt{s}$ $= 5.02$ TeV} 

\begin{figure}[htbp]
	\includegraphics[width=5cm]{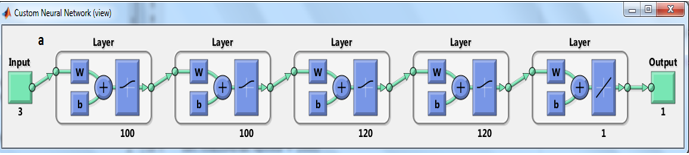}
	\includegraphics[width=5cm]{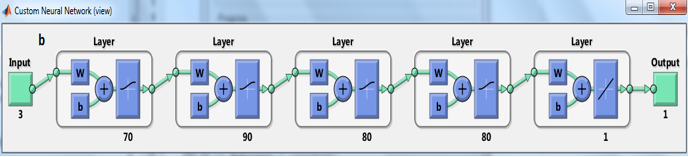}
	\includegraphics[width=5cm]{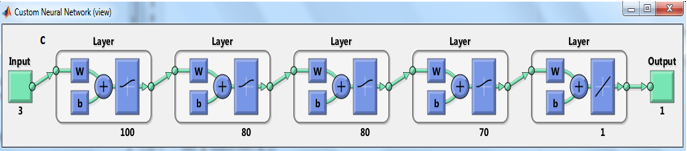}
	\includegraphics[width=5cm]{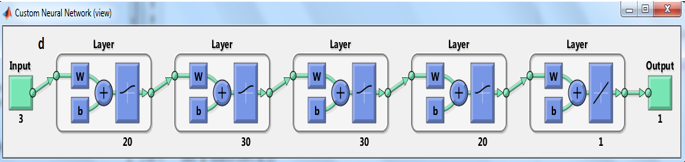}
	\includegraphics[width=5cm]{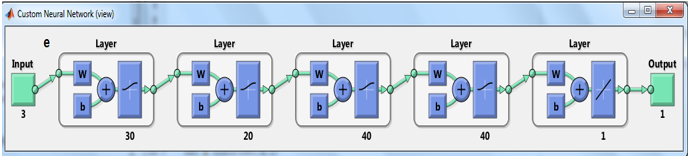}
	\caption{The same as in Fig. (\ref{fig:oneeai1}) but for particles (a) $\pi$, (b) $k$, (c) $p$, (d) $\Lambda$, and (e) $K_s^0$ at $\sqrt{s}$ $= 5.02$ TeV.}
	\label{fig:oneeai111}
\end{figure}

Here, In central Pb-Pb collisions at $\sqrt{s}$ $= 5.02$ TeV, we calculated the entropy per rapidity $d s/d y$ for particles $\pi$, $k$, $p$, $\Lambda$, and $K_s^0$. Transverse momentum spectra of the particles $\pi$, $k$, $p$ \cite{ALICE:2019hno}, $\Lambda$, and $K_s^0$ \cite{Sefcik:2018acn} are used as experimental input for the computation.
We also employ ALICE measured HBT source radii \cite{ALICE:2015hvw}. We also used the same deduced inputs for the ANN model. We applied the ANN model to acquire the $p_T$ spectra of the particles $\pi$, $k$, $p$, $\Lambda$, and $K_s^0$ according to the input parameters represented in Tab. (\ref{tabinputann502}). Five networks are chosen to simulate experimental data according to different particles. Best performance value and regression are obtained by using four hidden layers. The number of neurons in each hidden layer are ($100$, $100$, $120$, $120$), ($70$, $90$, $80$, $80$), ($100$, $80$, $80$, $70$), ($20$, $30$, $30$, $20$), ($30$, $20$, $40$, $40$) for particles $\pi$, $k$, $p$, $\Lambda$ and $K_s^0$, respectively. A simplification of the proposed ANN networks are shown in Fig.(\ref{fig:oneeai111}) for particles $\pi$(a), $k$(b), $p$(c), $\Lambda$(d), and $K_s^0$(e), respectively.

\begin{figure}[htbp]
\includegraphics[width=5cm]{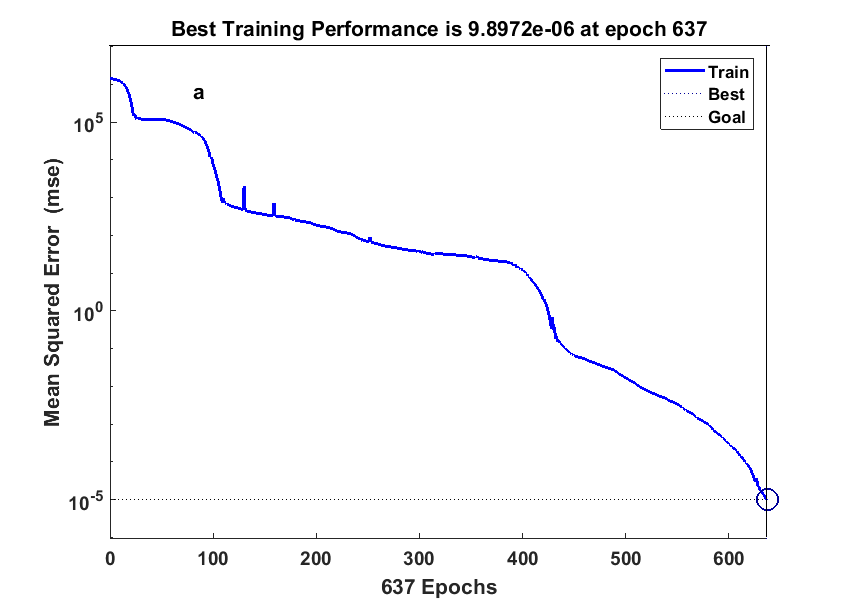}
\includegraphics[width=5cm]{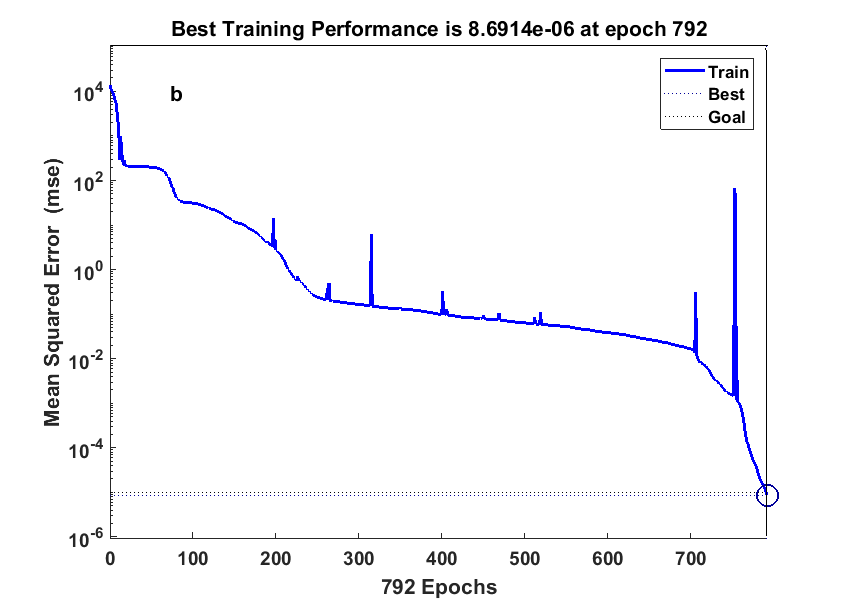}
\includegraphics[width=5cm]{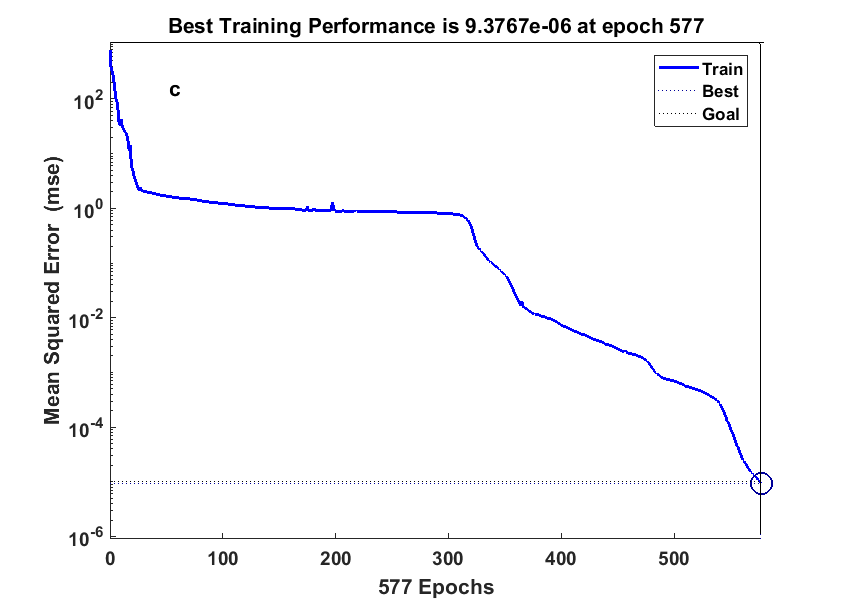}
\includegraphics[width=5cm]{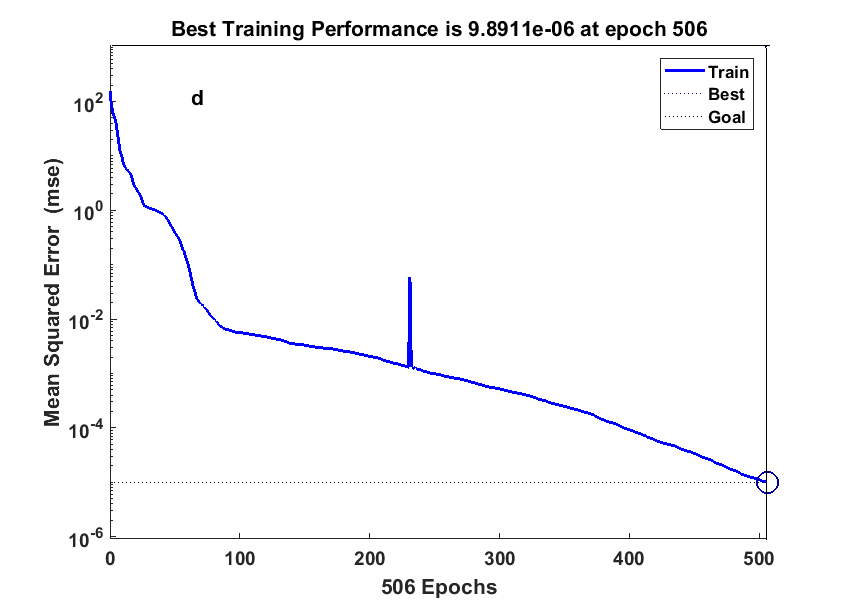}
\includegraphics[width=5cm]{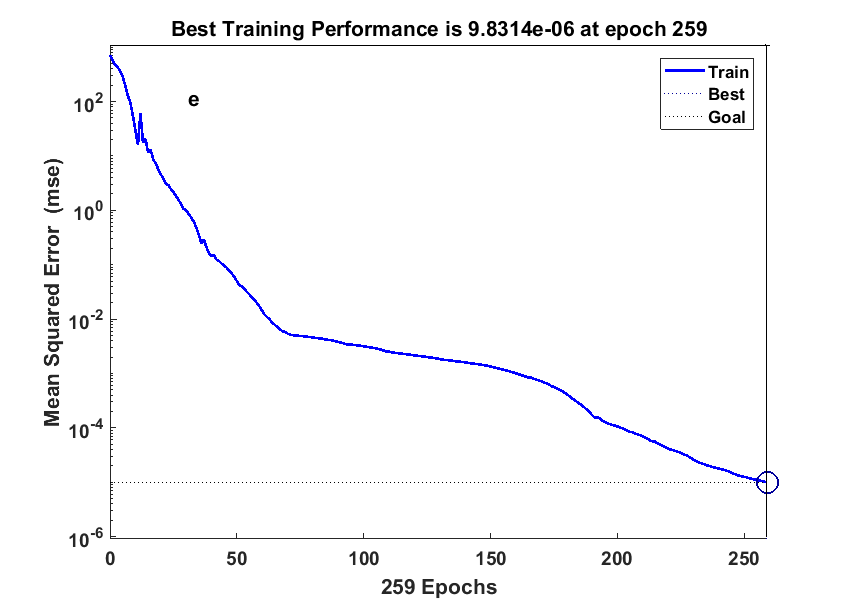}
\caption{The same as in Fig. (\ref{fig:oneeai2}) but for particles (a) $\pi$, (b) $k$, (c) $p$, (d) $\Lambda$, and (e) $K_s^0$ at $\sqrt{s}$ $= 5.02$ TeV.}
\label{fig:oneeai222}
\end{figure}

\begin{figure}[htbp]
\includegraphics[width=5cm]{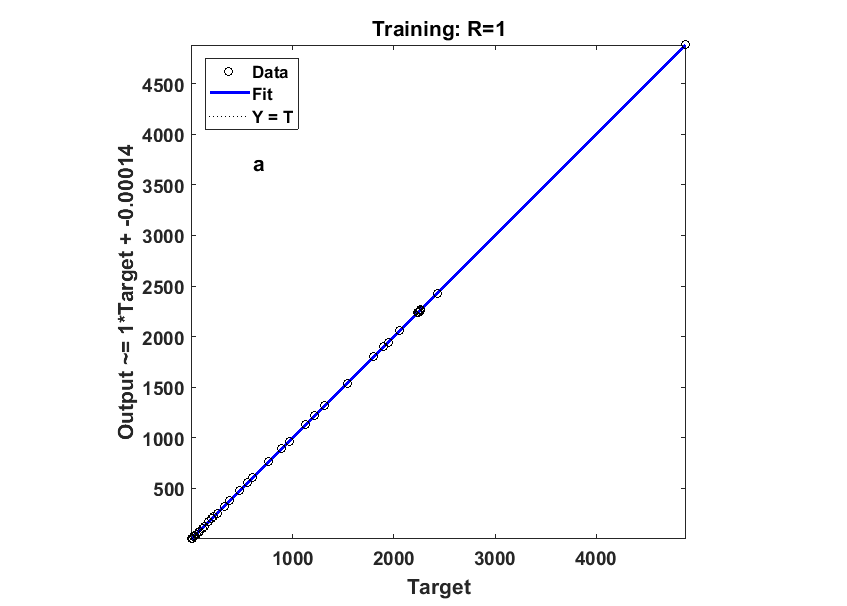}
\includegraphics[width=5cm]{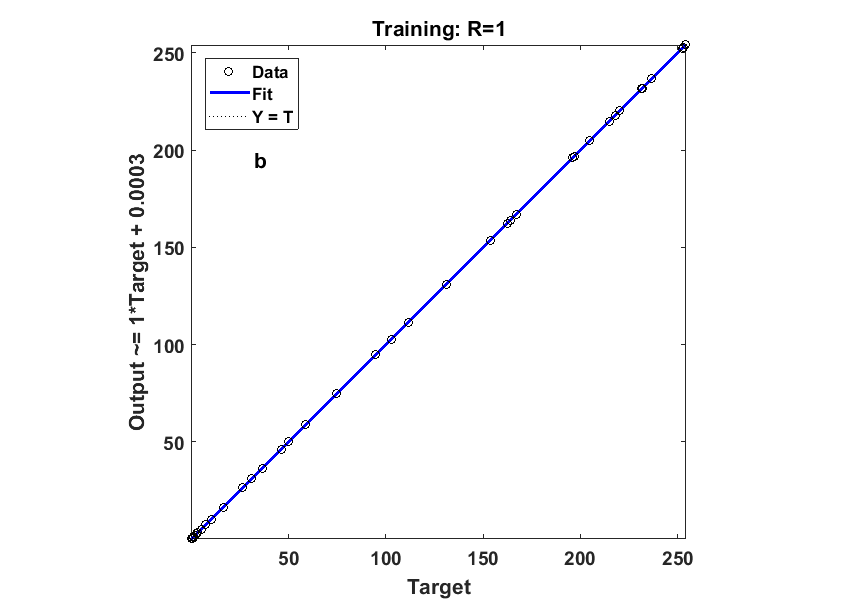}
\includegraphics[width=5cm]{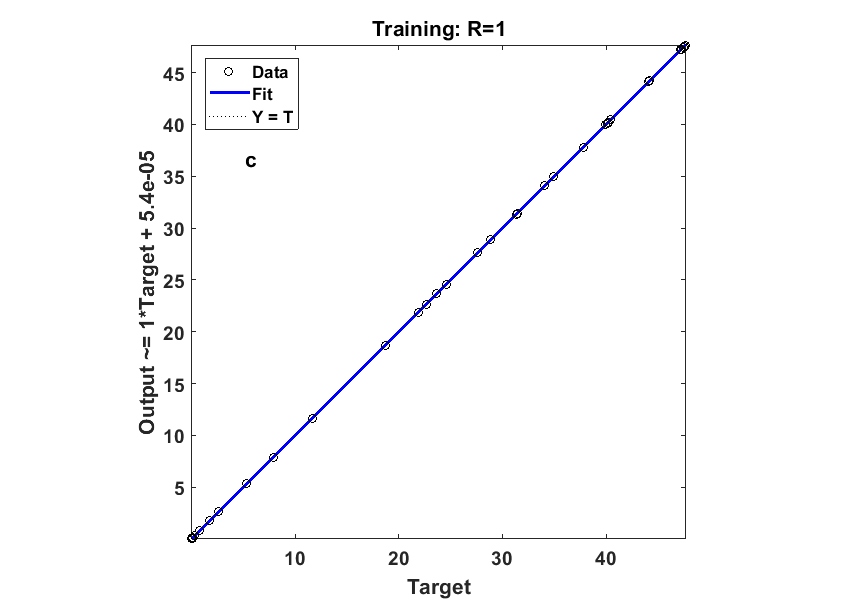}
\includegraphics[width=5cm]{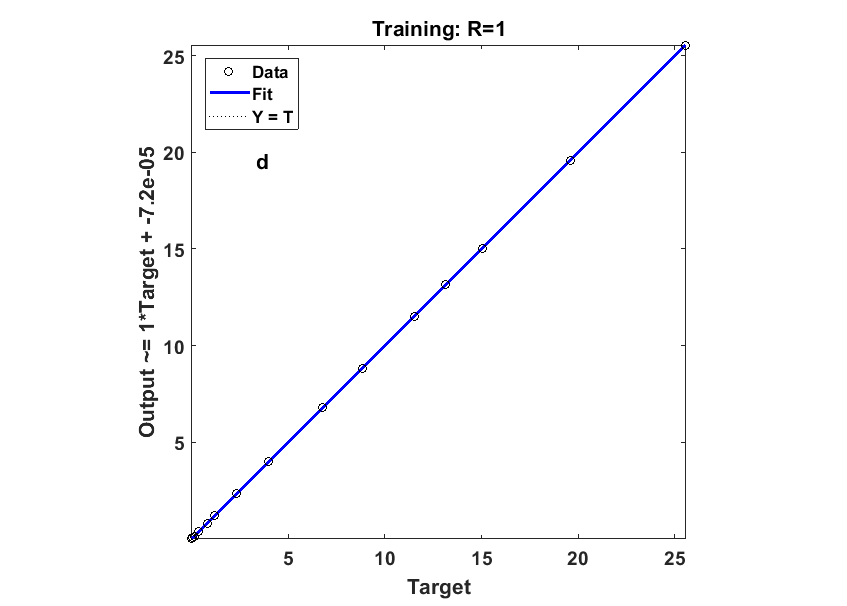}
\includegraphics[width=5cm]{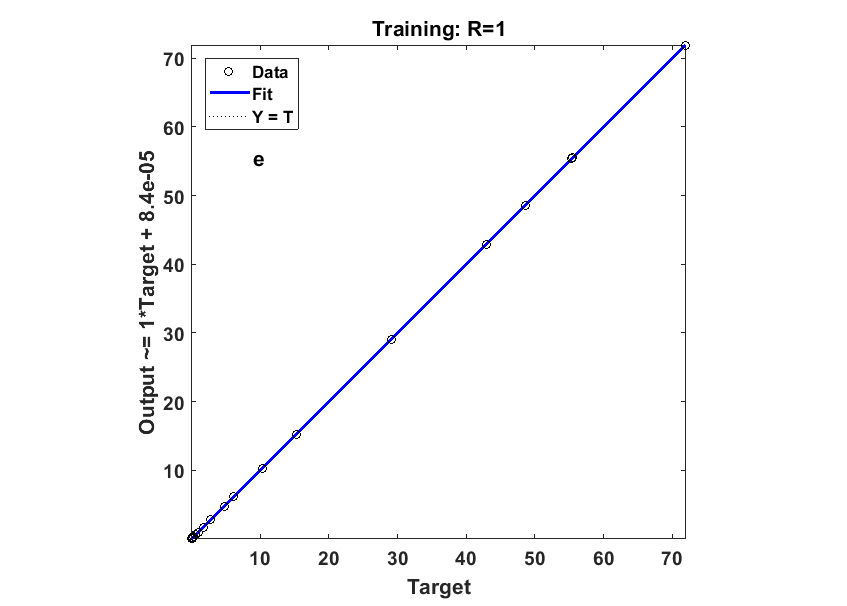}
\caption{The same as in Fig. (\ref{fig:oneeai3}) but for particles (a) $\pi$, (b) $k$, (c) $p$, (d) $\Lambda$, and (e) $K_s^0$ at $\sqrt{s}$ $= 5.02$ TeV.}
\label{fig:oneeai333}
\end{figure}

As a result, the obtained best performance and regression from training are shown in Figs.(\ref{fig:oneeai222} and \ref{fig:oneeai333}) for particles (a) $\pi$, (b) $k$, (c) $p$, (d) $\Lambda$, and (e) $K_s^0$ respectively. The performance is $9.897 \times 10^{-6}$, $8.6914 \times 10^{-6}$, $9.3767 \times 10^{-6}$, $9.8911 \times 10^{-6}$ and $9.8314 \times 10^{-6}$ after epoch $637$, $792$, $577$, $506$ and $259$ for particles  (a) $\pi$, (b) $k$, (c) $p$, (d) $\Lambda$, and (e) $K_s^0$ respectively as in Fig. (\ref{fig:oneeai222}). The transfer function used is \text{logsig} in hidden layers and \text{purelin} in output layer for all particles. All parameter used for ANN is shown in Tab.\ref{tabinputann502}. 

\begin {table}[htbp]
\caption {ANN parameters for particles $\pi$, $k$, $p$, $\Lambda$, and $K_s^0$ at $\sqrt{s}$ $= 5.02$ TeV.}
\begin{adjustbox}{width=\columnwidth}
\begin{tabular}{|c|c|c|c|c|c|}
	\hline
	ANN & \multicolumn{5}{|c|}{particles} \\
	\cline { 2 - 6 }
	parameters & $\pi$  & $K$  & $p$  & $\Lambda$ &$k_s^0$  \\
	\hline
	Inputs &  \multicolumn{2}{|c|}{$\sqrt{S} $}  &\multicolumn{2}{|c|}{$P_{T}(\mathrm{GeV})$} & Centrality \\
	\hline
	$\sqrt{S}$ &  \multicolumn{5}{|c|}{$5.02$ TeV}  \\
	\hline
	Output &  \multicolumn{5}{|c|}{$\frac{1}{N_{evt} }\frac{d^2 N}{ d y d p_T}$}   \\
	\hline
	Hidden layers & \multicolumn{5}{|c|}{4}  \\
	\hline
	Neurons & $100,100,120,120$ & $70,90,80,80$  &$100,80,80,70$  & $20,30,30,20$ &$30,20,40,40$  \\
	\hline
	Epochs & 637  &792  & 577 & 506 & 259\\
	\hline
	performance & $9.897 \times 10^{-6}$ & $8.6914 \times 10^{-6}$&$9.3767 \times 10^{-6}$  &$9.8911 \times 10^{-6}$  & $9.8314 \times 10^{-6}$  \\
	\hline
	Training algorithms & \multicolumn{5}{|c|}{Rprop}   \\
	\hline
	Training functions & \multicolumn{5}{|c|}{trainrp}   \\
	\hline
	Transfer functions of hidden layers & Logsig & Logsig & Logsig & Logsig & Logsig \\
	\hline
	Output functions & \multicolumn{5}{|c|}{Purelin} \\
	\hline
\end{tabular}
\end{adjustbox}
\label{tabinputann502}
\end {table}

Extrapolation of the observed transverse momentum spectra to $p_T = 0$ is necessary to determine the entropy $S$. To achieve this, we fitted both the experimental and simulated $p_T$ spectra to two various functional models, Tsallis distribution \cite{Cleymans:2016opp,Bhattacharyya:2017hdc} and the HRG model \cite{Yassin:2019xxl}. The aim of combining two models is to fit the entire $p_T$ curve. 

\begin{figure}[htbp]
\includegraphics[width=5cm]{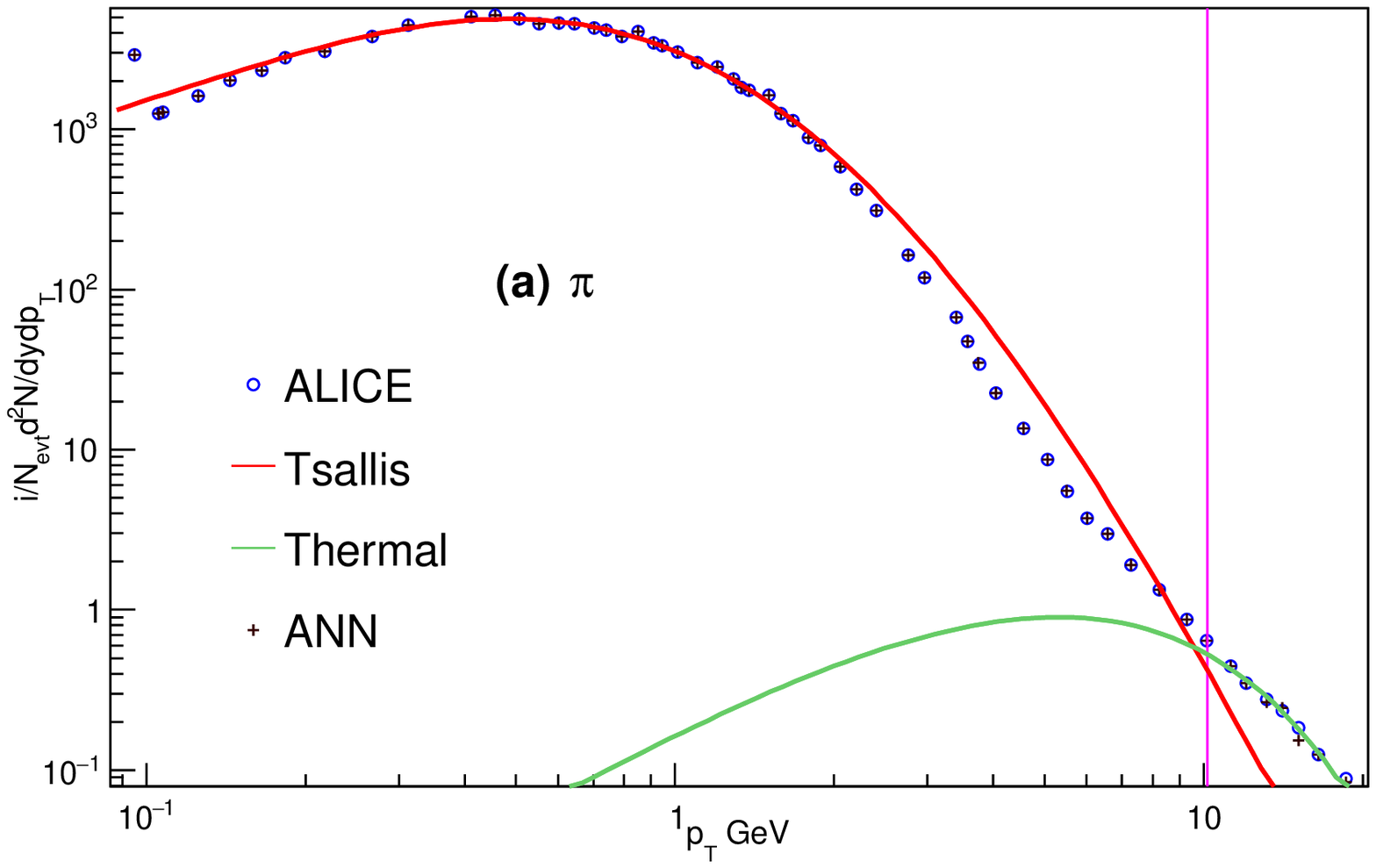}
\includegraphics[width=5cm]{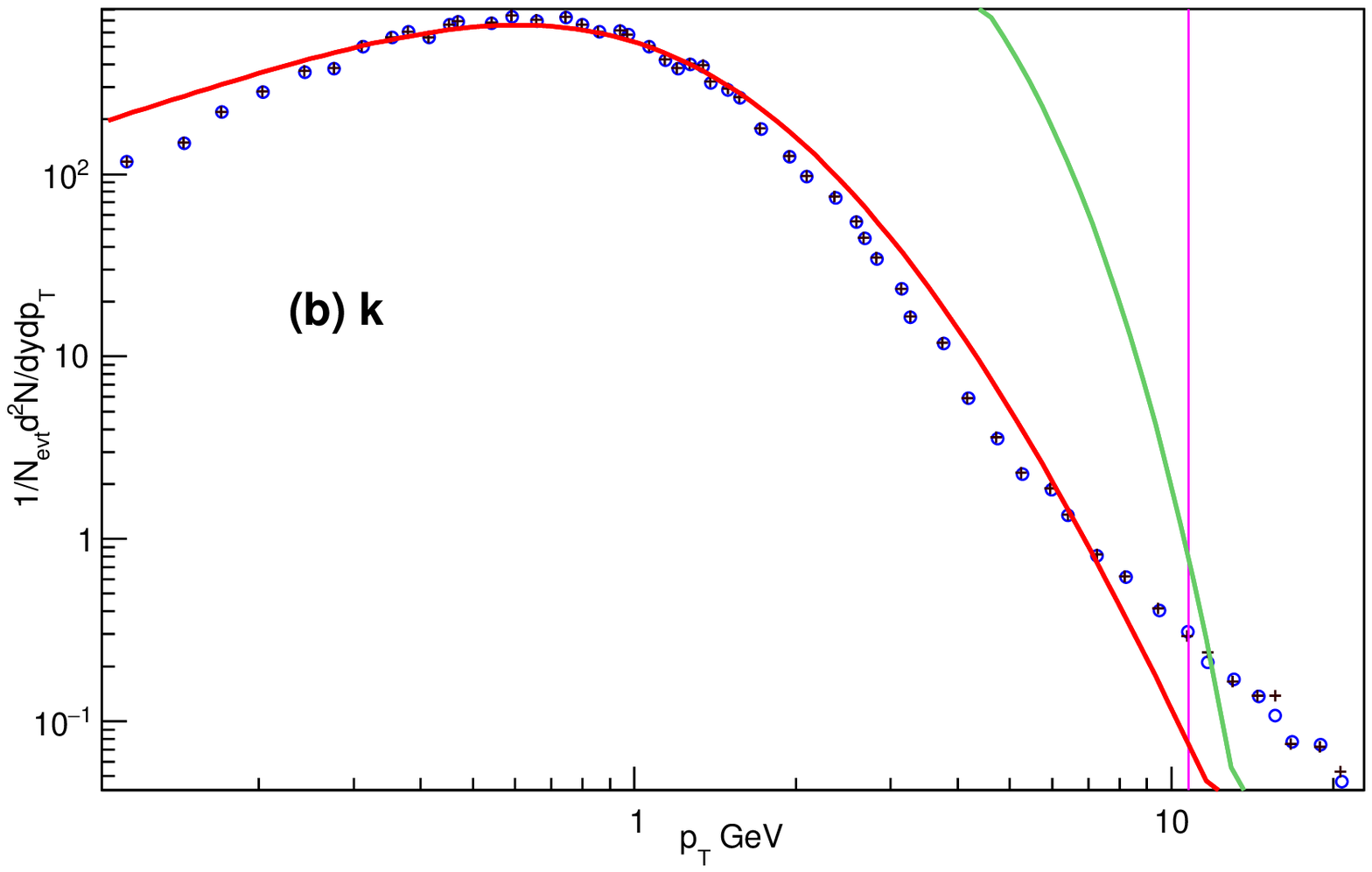}
\includegraphics[width=5cm]{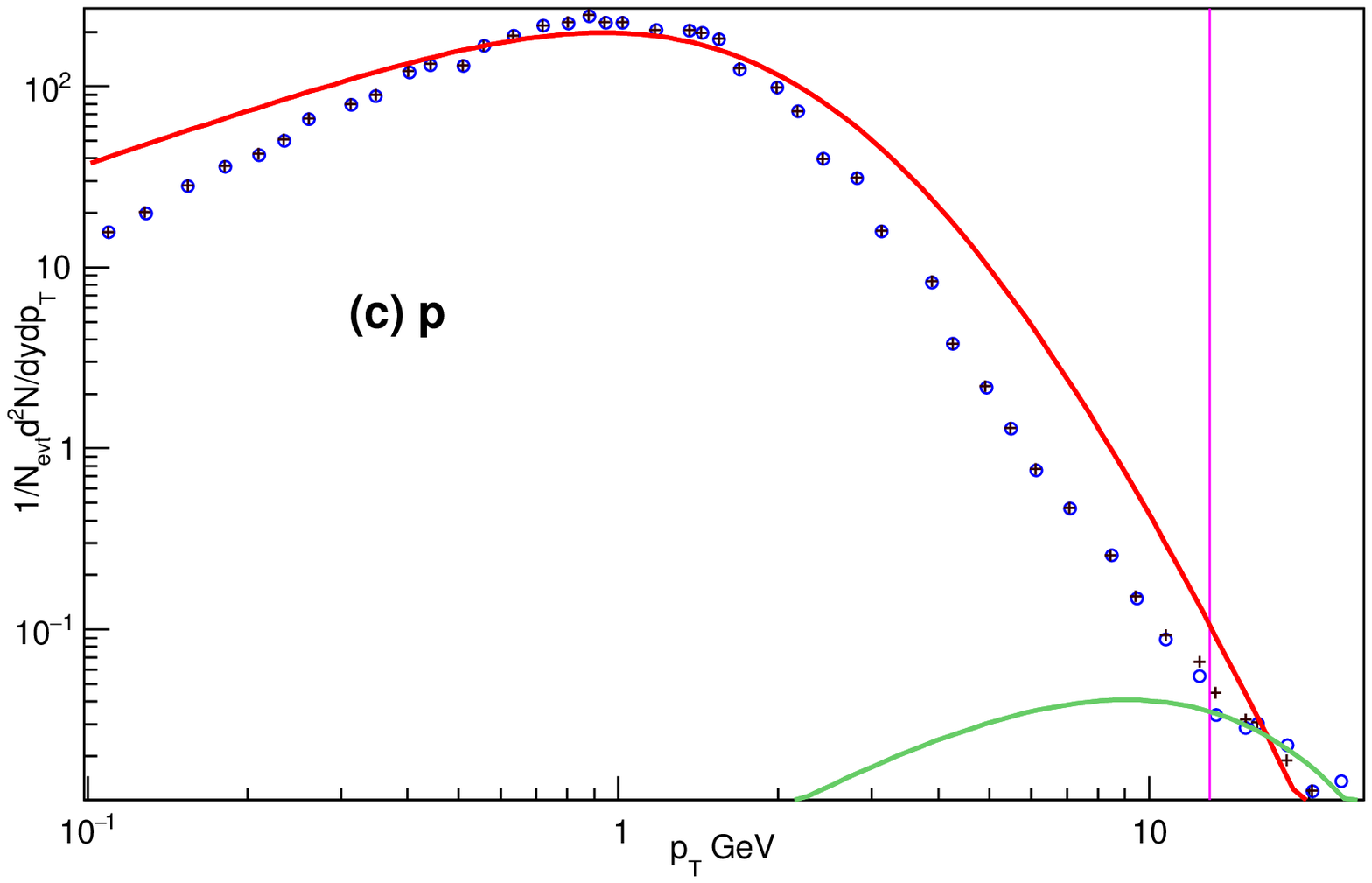}
\includegraphics[width=5cm]{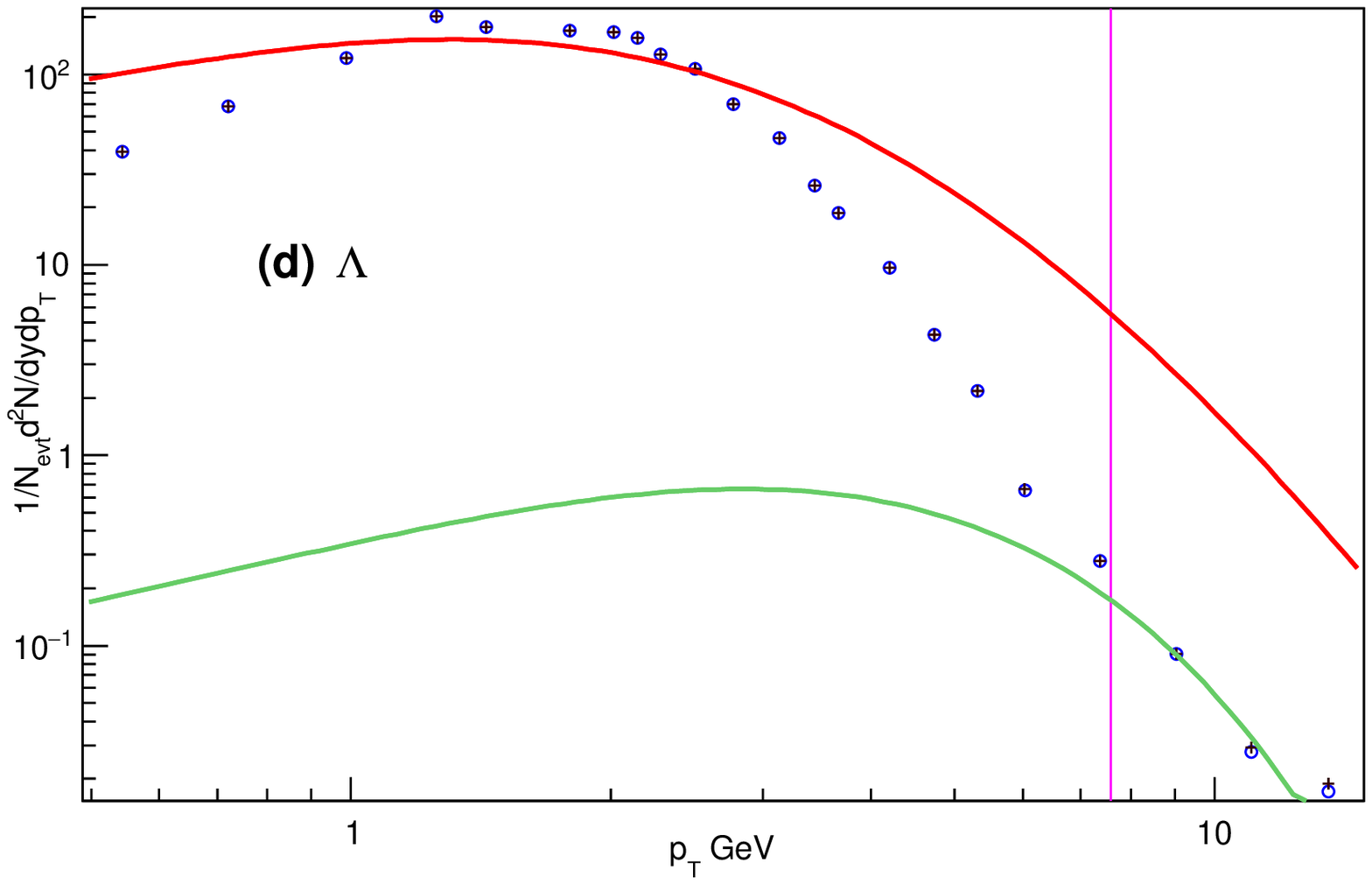}
\includegraphics[width=5cm]{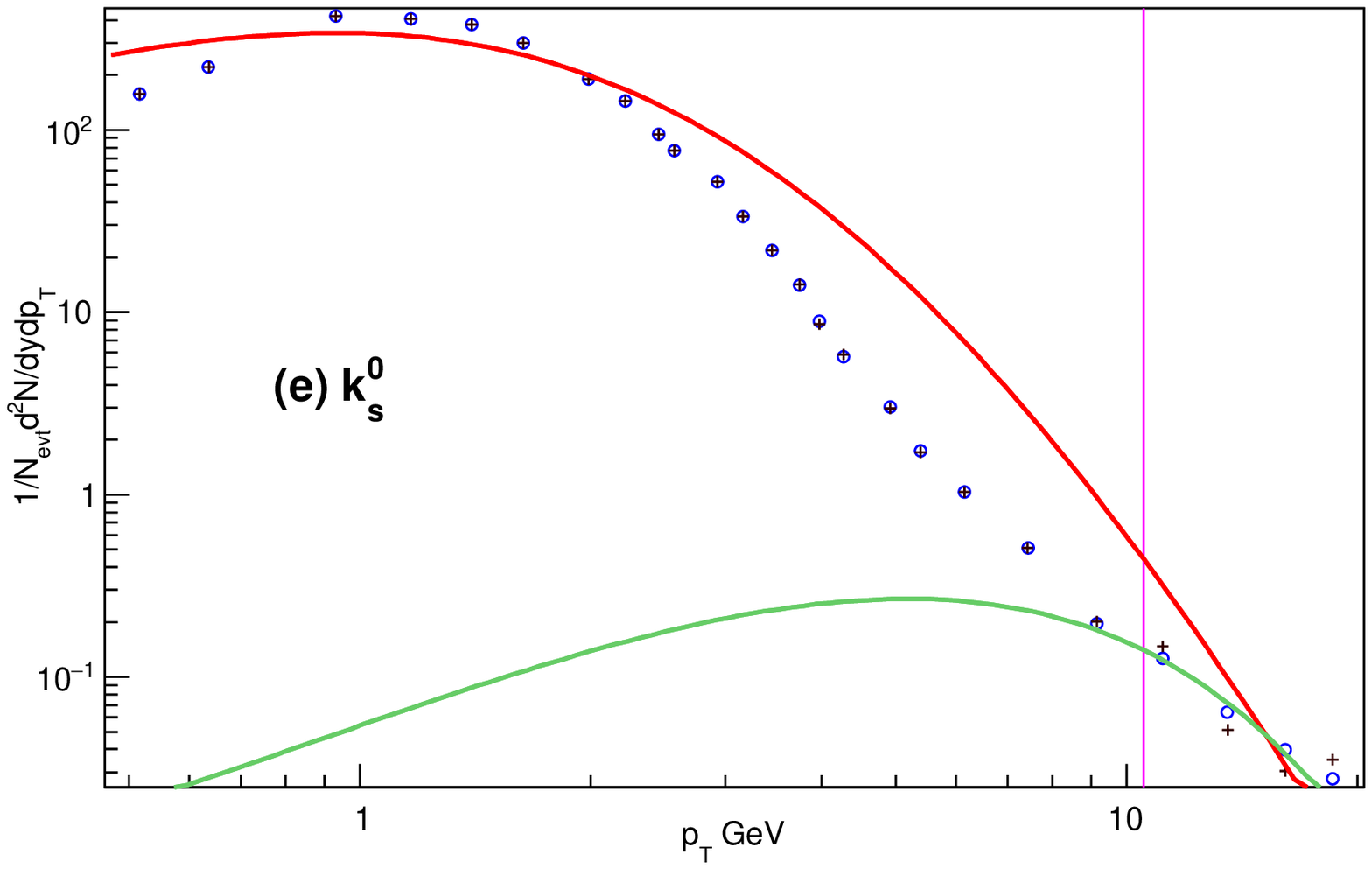}
\caption{The transverse momentum distribution, measured by ALICE experiment collaboration \cite{ALICE:2019hno,Sefcik:2018acn} at centre of mass energy $= 5.02$ TeV and represented by blue open circles symbols, for particles $\pi$ (a), $k$ (b), $p$ (c), $\Lambda$ (d), and $K_s^0$ (e) is compared to the statistical fits from Tsallis distribution, perfectly fits at $0.001 < y < 10 $, represented by red solid line given by Eq. (\ref{eqb:5}) and the HRG model, works in $ 10 < y < 20 $, represented by solid green line given by Eq. (\ref{eqq(7)}). A boarder line is drawn between Tsallis and HRG models and represented by solid purple color. The experimental data and the results of both models are then confronted to that obtained from the ANN simulation model, represented by dark brown plus sign symbols.}  
\label{fig:twonnn}
\end{figure}

In Fig. (\ref{fig:twonnn}), the experimental and simulated particle spectra $p_T$ are compared to the Tsallis distribution and the HRG model for particles $\pi$ (a), $k$ (b), $p$ (c), $\Lambda$ (d), $\Omega$ (e), and $\bar{\Sigma}$ (f). As seen in Fig. (\ref{fig:twon}), employing various forms of the fitting function is obvious, as the Tsallis function can only match the left side of the $p_T$ curve at $0.001 < y < 10 $, whereas the HRG model can fit the right side at  $ 10 < y < 20 $. This result may motivate us to pursue additional research. Tabs. (\ref{tab1exfitt502}) and (\ref{tab1annfitt502}) summarise the fitting parameters obtained from the Tsallis distribution and HRG model, respectively.

\begin {table}[htbp]
\caption {The transverse momentum distribution fitting parameters when confronting the Tasllis distribution, Eq. (\ref{eqq(7)}), and the HRG model, Eq. (\ref{eqb:5}), to the ALICE experiment data \cite{ALICE:2019hno,Sefcik:2018acn} at $\sqrt{s}$ $= 5.02$ TeV for particles $\pi$, $k$, $p$, $\Lambda$, and $K_s^0$.}
\begin{adjustbox}{width=\columnwidth}
\begin{tabular}{|c|c|c|c|c|c|c|c|}
\hline 
 \multirow{2}{*}{particle} & \multicolumn{3}{c|}{Tsallis parameters}&\multicolumn{3}{c|}{HRG model} & \multirow{2}{*}{$\chi^{2}$ /dof} \\ 
 \cline{2-7}
  & $ dN/dy$  & $T$ GeV & q &V $fm^3$  &$T$ GeV & $\mu$ GeV &  \\
\hline 
 $\pi$& $5804.25$ & $0.2101$ & $1.1218$ & $4709.14 \pm 1351.88$ &$ 2.6497 \pm 0.0475$& $26.2268 \pm 0.7773$&   $316/48$ \\ 
\hline 
   $K$ &$ 953.214$ &$ 0.2392$ &$1.1146$ & $3614.78 \pm 1.6854$ &$0.719156 \pm 630181$ & $1.28019 \pm 630181$&  $905.7/44  $ \\ 
\hline 
 $p$ & $437.394$  &$0.3393$ & $1.1167$ & $ 8.6667 \pm 3.7659$&$3.45715 \pm 0.1372$ &$22.842 \pm 1.8419$ & $219.6/36 $   \\ 
\hline
  $\Lambda$  & $466.969$ & $0.4920$ &$1.119$  & $395.622 \pm 44731.7$&$1.49141 \pm 15.7463$ &$ 10.0101 \pm 226.948$& $198/16$ \\ 
\hline
  $K_s^0$ &$750.684$ &$ 0.4003$ & $1.105$ & $72.7633 \pm 82.7046$&$2.5859 \pm 0.227$ &$17.809 \pm 3.2296$ &  $547.7/18$     \\ 
\hline  
\end{tabular} 
\end{adjustbox}
\label{tab1exfitt502}
\end {table}

\begin {table}[htbp]
\caption {The same in Tab. (\ref{tab1exfitt502}) but the statistical fits results, from both used models, is confronted to the ANN simulation model.}
\begin{adjustbox}{width=\columnwidth}
\begin{tabular}{|c|c|c|c|c|c|c|c|}
\hline 
 \multirow{2}{*}{particle} & \multicolumn{3}{c|}{Tsallis fitting parameters}&\multicolumn{3}{c|}{HRG model fitting parameters} & \multirow{2}{*}{$\chi^{2}$ /dof} \\ 
 \cline{2-7}
  &  $dN/dy$  & $T_{Ts}$ GeV & q &V $fm^3$  &$T_{th}$ GeV & $\mu$ GeV &  \\
\hline 
 $\pi$& $5807.58$ & $0.2099$ & $1.1221$ & $4681.65 \pm 1995.89$ &$ 2.5486 \pm 0.066$&$ 24.7643 \pm 1.0789$&$3173/48$   \\ 
\hline 
   $K$ &$958.826$ & $0.2396$& $1.1159$& $ 5871.87 \pm 4273.75$ &$ 3.202 \pm 0.1495$ & $36.6246 \pm 2.627$ &$9116.2/44    $ \\ 
\hline 
 $p$ & $440.362$ & $ 0.3341$& $1.1223$ &$1981.5 \pm 1398.83$ &$ 4.5297 \pm 0.2092$ &$59.7687 \pm 3.7575$ & $2246.4/36   $  \\ 
\hline
  $\Lambda$  &$472.765$ & $0.4976$ &$ 1.1183 $& $396.77 \pm 53486.2$ &$1.42152 \pm 15.2713$ &$ 9.10505 \pm 220.834$ & $  1949.7/16$ \\ 
\hline
  $K_s^0$ &$749.794 $ & $ 0.3945$& $1.108$ &$ 26.688 \pm 55.3559$ &$2.0758 \pm 0.4493$ &$9.7132 \pm 5.5569$ &     $5558.9/18 $\\ 
\hline  
\end{tabular} 
\end{adjustbox}
\label{tab1annfitt502}
\end {table}  
   
The estimated entropy per rapidity $ds/dy$ from Pb-Pb central collisions at $\sqrt{s}$ $= 5.02$ TeV using the Tsallis distribution, HRG model, and  ANN model for particles $\pi$, $k$, $p$, $\Lambda$, and $K_s^0$ is represented in Tab. (\ref{tab3502entropy}). The effect of both the Tsallis distribution and HRG model fitting function on the estimated entropy per rapidity $d s/d y$ is also shown in Tab. (\ref{tab3502entropy}). 
 
\begin {table}[htbp]
\caption {The same in Tab. (\ref{tab3276entropy}) but at $\sqrt{s}$ $=5.02$ TeV.}
\begin{adjustbox}{width=\columnwidth}
\begin{tabular}{|c|c|c|c|c|}
\hline 
 particle & $(ds/dy)_{y=0}$ & $(ds/dy)_{y=0}$ supplemented by Tsallis& $(ds/dy)_{y=0}$ supplemented by HRG model & $(ds/dy)_{y=0}$ estimated by ANN model \\ 
\hline 
$\pi$ &  $13188.2$ & $13406.9$ &$13333.6$   & $13330.9$    \\ 
\hline 
  $K$ & $3909.58 $& $3938.4 $&$1011.8  $&$3896.25 $  \\ 
\hline 
 $p$&  $2259.36$ &$ 2262.64$ & $2250.08$ & $2253.3$     \\ 
\hline
  $\Lambda$ &$2104.91 $  &$2193.98 $ & $2179.42 $&$2178$     \\ 
\hline
 $K_s^0$ & $2463.9$  &$3396.13$  & $3372.08$ & $3369.06$   \\ 
\hline  
\end{tabular} 
\end{adjustbox}
\label{tab3502entropy}
\end {table}

 The values of the entropy per rapidity $d s/d y$ are calculated from fitting the experimental and simulated particle spectra to the statistical models are agree with each other. The function which describes the non-linear relationship between inputs and output is given in Appendix \ref{sec:(append:neural)}. This implies further use for ANN model to predict the entropy per rapidity $d s/d y$ in the absence of the experiment.

\section{Summary and Conclusions}
\label{sec:Cncls}

In this work, We calculated the entropy per rapidity $d S/d y$ produced in central Pb-Pb ultra-relativistic nuclear collisions at LHC energies using experimentally observed identifiable particle spectra and source radii estimated from HBT correlations. The considered particles are $\pi$, $k$, $p$, $\Lambda$, $\Omega$, and $\bar{\Sigma}$, and $\pi$, $k$, $p$, $\Lambda$, and $K_s^0$ where the center of mass energy is $ \sqrt{s}$ $=2.76$ and $5.02$ TeV, respectively.  ANN simulation model is used to estimate the entropy per rapidity $d S/d y$ for the same particles at the considered energies. Extrapolating the transverse momentum spectra at $p_T$ $=0$ is required to calculate $d S/d y$ thus we use two different fitting functions, Tsallis distribution and the Hadron Resonance Gas (HRG) model. The effect of both the Tsallis distribution and HRG model fitting function on the estimated entropy per rapidity $d s/d y$ is also discussed. The Tsallis function can only match the left side of the $p_T$ curve, whereas the HRG model can fit the right side. This result may motivate us to pursue additional research. The success of ANN model to describe the experimental measurements will implies further prediction for the entropy per rapidity in the absence of the experiment.

\begin{appendices}
\label{appendces}
\section{A detailed description for the entropy production $d s/ d y$ as shown in Eq. (\ref{eq:1})}
\label{sec:(append:entropy)}
\renewcommand{\theequation}{A.\arabic{equation}}
\setcounter{equation}{0}

According to Gibbs-Duham relation, the thermodynamic quantities are related by \cite{Letessier:2002gp}

\begin{equation}\label{eq:s1}
E(V, T, \boldsymbol{\mu})=\boldsymbol{F}^{\prime}(V, T, \boldsymbol{\mu})+T S(V, T, \boldsymbol{\mu})+\boldsymbol{\mu} \boldsymbol{b}(V, T, \boldsymbol{\mu})
\end{equation}

Thus the entropy $(S)$ can be obtained as\cite{Letessier:2002gp}
\begin{equation}\label{eq:s2}
S=\frac{1}{T}\left(E-\boldsymbol{F}^{\prime}-\boldsymbol{\mu} \boldsymbol{b}\right)=\ln Z-\beta \frac{\partial \ln Z}{\partial \beta}-(\ln \lambda) \boldsymbol{\lambda} \frac{\partial \ln Z}{\partial \lambda}
\end{equation}

Our aim is to write $(S)$ in terms of $(f)$, which is the single particle distribution function, and it is given by\cite{Letessier:2002gp}
\begin{equation}\label{eq:s3}
\mathrm{f}_{F / \beta}(\xi, \boldsymbol{\beta}, \lambda)=\frac{1}{e^{\beta(\xi-\mu)} \pm \mathbf{1}}
\end{equation}
where $(+)$ and $(-)$ represent fermions and bosons, respectively.

The partition function ( $\ln Z $) is given by\cite{Letessier:2002gp}
\begin{equation}\label{eq:s4}
\ln Z_{F / \beta}(\boldsymbol{V}, \boldsymbol{\beta}, \lambda)=\pm \int \frac{d^{3} r d^{3} P}{(2 \pi)^{3}} \ln \left[1 \pm \boldsymbol{e}^{\beta(\mu-\xi)}\right]
\end{equation}

Differentiating Eq. (\ref{eq:s4}) with respect to $\beta$, the inverse of temperature, we get \cite{Letessier:2002gp}

\begin{equation}\label{eq:s5}
\begin{array}{l}
\frac{\partial \ln Z_{F / \beta}}{\partial \beta}=\pm \int \frac{d^{3} r d^{3} P}{(2 \pi)^{3}} \frac{(\boldsymbol{\mu}-\boldsymbol{\xi}) \boldsymbol{e}^{\boldsymbol{\beta}(\mu-\xi)}}{1 \pm \boldsymbol{e}^{\boldsymbol{\beta}(\mu-\xi)}}\\
=\pm \int \frac{d^{3} r d^{3} P}{(2 \pi)^{3}} \frac{(\mu-\xi)}{e^{\beta(\xi-\mu)} \pm 1}
\end{array}
\end{equation}

Also, Differentiating Eq. (\ref{eq:s4}) with respect to $\lambda$, we get \cite{Letessier:2002gp} 

\begin{equation}\label{eq:s6}
\frac{\partial \ln Z}{\partial \lambda}=\pm \int \frac{d^{3} r d^{3} P}{(2 \pi)^{3}} \frac{\boldsymbol{e}^{\boldsymbol{\beta}(\mu-\xi)}}{1 \pm \boldsymbol{e}^{\boldsymbol{\beta}(\boldsymbol{\mu}-\xi)}}
\end{equation}

Eq. (\ref{eq:s6}) can be arranged as \cite{Letessier:2002gp}
\begin{equation}\label{eq:s7}
\frac{\partial \ln Z}{\partial \lambda}=\pm \int \frac{d^{3} r d^{3} P}{(2 \pi)^{3}} \frac{1}{e^{\beta(\xi-\mu)} \pm 1}
\end{equation}

Substituting from Eqs. (\ref{eq:s4}), (\ref{eq:s5}), and (\ref{eq:s7}) into Eq. (\ref{eq:s2}), we get \cite{Letessier:2002gp}
\begin{equation}\label{eq:s8}
\begin{array}{c}
S=\pm \int \frac{d^{3} r d^{3} P}{(2 \pi)^{3}}\left[\ln \left(1 \pm e^{\beta(\mu-\xi)}\right)-\frac{\beta(\mu-\xi)}{e^{\beta(\mu-\xi)} \pm 1}\right. \\
\left.-\frac{\beta \mu e^{\beta \mu}}{e^{\beta(\xi-\mu)} \pm \mathbf{1}}\right]
\end{array}
\end{equation}

at vanishing chemical potential, $\mu_B=0 $, the last term in Eq. (\ref{eq:s8}) will be equal zero.

\begin{equation}\label{eq:s9}
\boldsymbol{e}^{\boldsymbol{\beta}(\xi-\boldsymbol{\mu})} \pm \mathbf{1}=\frac{1}{\mathrm{f}_{F / \beta}}
\end{equation}

Eq. (\ref{eq:s3}) can be written in the following form \cite{Letessier:2002gp}
\begin{equation}\label{eq:s10}
-e^{\beta(\xi-\mu)}=\frac{1}{\mathrm{f}_{F / \beta}} \mp 1=\frac{1 \pm \mathrm{f}_{F / \beta}}{\mathrm{f}_{F / \beta}}
\end{equation}

recalling Eq. (\ref{eq:s10}), we obtain
\begin{equation}\label{eq:s11}
-e^{\boldsymbol{\beta}(\boldsymbol{\mu}-\xi)}=\frac{\mathrm{f}_{F / \beta}}{1 \mp \mathrm{f}_{F / \beta}}
\end{equation}

Rearranging Eq. (\ref{eq:s11}) in the following form
\begin{equation}\label{eq:s12}
1 \pm e^{\beta(\mu-\xi)}=1 \pm \frac{\mathrm{f}_{F / \beta}}{1 \mp \mathrm{f}_{F / \beta}}=\frac{1 \mp \mathrm{f}_{F / \beta} \pm \mathrm{f}_{F / \beta}}{1 \mp \mathrm{f}_{F / \beta}}-\frac{1}{1 \mp \mathrm{f}_{F / \beta}}=1 \pm e^{\beta(\mu-\xi)}
\end{equation}

Substituting from Eq. (\ref{eq:s12}) into Eq. (\ref{eq:s8}), we get 
\begin{equation}\label{eq:s13}
S=\pm \int \frac{d^{3} r d^{3} P}{(2 \pi)^{3}}\left[\ln \left(\frac{1}{1 \mp \mathrm{f}_{F / \beta}}\right)-\ln \left(\frac{\mathrm{f}_{F / \beta}}{1 \mp \mathrm{f}_{F / \beta}}\right) \mathrm{f}_{F / \beta}-z e r o\right]
\end{equation}

Rearranging Eq. (\ref{eq:s13}) as 
\begin{equation}\label{eq:s14}
S=\pm \int \frac{d^{3} r d^{3} P}{(2 \pi)^{3}}\left[-\ln \left(1 \mp \mathrm{f}_{F / \beta}\right)-\left[\ln \mathrm{f}_{F / \beta}-\ln \left(1 \mp \mathrm{f}_{F / \beta}\right)\right] \mathrm{f}_{F / \beta}\right]
\end{equation}

Simplifying Eq. (\ref{eq:s14}) as 
\begin{equation}\label{eq:s15}
\begin{array}{c}
S=\pm \int \frac{d^{3} r d^{3} P}{(2 \pi)^{3}}\left[-\ln \left(1 \mp \mathrm{f}_{F / \beta}\right)-\mathrm{f}_{F / \beta} \ln \mathrm{f}_{F / \beta}\right. \\
\left.+\mathrm{f}_{F / \beta} \ln \left(1 \mp \mathrm{f}_{F / \beta}\right)\right]
\end{array}
\end{equation}

Finally, the entropy $S$ can be given by \cite{Letessier:2002gp}
 
\begin{equation}\label{eq:s16}
\begin{array}{c}
S=\pm \int \frac{d^{3} r d^{3} P}{(2 \pi)^{3}}\left[-\mathrm{f}_{F / \beta} \ln \left(1 \mp \mathrm{f}_{F / \beta}\right)-\mathrm{f}_{F / \beta} \ln \mathrm{f}_{F / \beta}\right. \\
\left.+\mathrm{f}_{F / \beta} \ln \left(1 \mp \mathrm{f}_{F / \beta}\right)\right].
\end{array}
\end{equation}

Eq. (\ref{eq:s16}) represents the entropy equation that shown in Eq. (\ref{eq:1}).

\section{The transverse momentum distribution based on the HRG model}
\label{sec:(append:hrg)}
\renewcommand{\theequation}{B.\arabic{equation}}
\setcounter{equation}{0}

 The partition function $Z(T,V,\mu)$ is given by 

\begin{equation} 
Z(T,V,\mu)=\mbox{Tr}\left[\exp\left(\frac{{\mu}N-H}{\mathtt{T}}\right)\right],
\label{eqq(1)}
\end{equation}
where $H$ stands for the system's Hamiltonian, $\mu$ is the chemical potential, and  $N$ is the net number of all constituents. In the
HRG approach, Eq. (\ref{eqq(1)}) can be written as a summation of all hadron resonances 
\begin{equation}\ln Z(T,V,\mu)=\sum_i{{\ln Z}_i(T,V,\mu)} =\frac{V g_i}{(2{\pi})^3}\int^{\infty}_0{\pm 
d^{3}p {\ln} {\left[1\pm \exp\left(\frac{E-\mu_{i}}{\mathtt{T}} \right)
\right]}}, \label{eqq(2)}
\end{equation}

where $\pm $ represent the bosons and fermions particles, respectively and
$E_{i}=\left(p^{2}+m_{i}^{2}\right)^{1/2}$ is the energy of the $i$-th hadron.

the particle's multiplicity can be determined from the partition function as 
\begin{equation}
N_{i}=T \frac{\partial Z_{i}(T, V)}{\partial \mu_{i}}=\frac{V g_i}
{(2{\pi})^3}\int^{\infty}_0{d^{3}p \left[\exp\left(\frac{E-\mu_{i}}
{\mathtt{T}}\right)\pm1\right]^{-1}}, \label{eqq(3)} 
\end{equation}

For a partially radiated thermal source, the inavriant momentum spectrum is obtained as \cite{Letessier:2002gp}

\begin{equation}
 E\frac{d^{3}N_{i}}{d^{3}p}=E \frac{V g_i}
{(2{\pi})^3}\left[\exp\left(\frac{E-\mu_{i}}
{\mathtt{T}}\right)\pm1\right]^{-1}, \label{eqq(4)}
\end{equation}

The i-th particle's energy $E_{i}$ can be written as a function of the rapidity $\left(y\right)$ and $m_{T}$ as
\begin{equation}
E=m_{T} \cosh\left(y\right), \label{eqq(5)}
\end{equation}

Where $m_{T}$ represents the transverse mass and can be written in terms of the transverse momentum $p_{T}$ by 
\begin{equation}
m_{T}=\sqrt{m^{2}+p_{T}^{2}}, \label{eqq(6)}
\end{equation}

Substituting from Eq.(\ref{eqq(5)}) in to Eq.(\ref{eqq(4)}), one get the particle momentum distribution

at mid-rapidity ($y= 0$) and $\mu \neq 0$ 

\begin{equation}
\frac{1}{2 \pi p_{T}} \frac{d^2N}{dydp_{T}}= \frac{V g_i m_{T}}
{(2{\pi})^3}\left[\exp\left(\frac{m_{T}-\mu_{i}}
{\mathtt{T}}\right)\pm1\right]^{-1}. \label{eqq(7)}
\end{equation}

We fitted the experimental data of the particle momentum spectra with that calculated from Eq.(\ref{eqq(7)}) where the fitting parameters are $V$, $\mu$, and $\mathtt{T}$.

\section{The transverse momentum distribution based on Tsallis model}
\label{sec:(append:tsalis)}
\renewcommand{\theequation}{C.\arabic{equation}}
\setcounter{equation}{0}

 The transverse momentum distribution of the produced hadrons at LHC energies \cite{Cleymans:2016opp,Bhattacharyya:2017hdc}

\begin{equation} 
\left.\frac{1}{p_{T}} \frac{d^{2} N}{d p_{T} d y}\right|_{y=0}=g V \frac{m_{T}}{(2 \pi)^{2}}\left[1+(q-1) \frac{m_{T}}{T}\right]^{-q /(q-1)}, \label{eqb:1}
\end{equation}

where $m_{T}$ and $p_{T}$ represent the transverse mass and transverse momentum, respectively. $y$ is the rapidity, $g$ is the degeneracy factor, and $V$ is the volume of the system. 

The obtained values of $q$ and $T$ represent a system in the kinetic freeze-out case.

In the limit where $q->1$, Eq. (\ref{eqb:1}) is a simplification of the conventional Boltzmann distribution as \cite{Cleymans:2016opp,Bhattacharyya:2017hdc}

\begin{equation} 
\left.\lim _{q \rightarrow 1} \frac{1}{p_{T}} \frac{d^{2} N}{d p_{T} d y}\right|_{y=0}=g V \frac{m_{T}}{(2 \pi)^{2}} \exp \left(-\frac{m_{T}}{T}\right), \label{eqb:2}
\end{equation}

As a result, several statistical mechanics ideas may be applied to the distribution provided in Eq. (\ref{eqb:1}).

Integrating Eq. (\ref{eqb:1}) though the transverse momentum, one gets \cite{Cleymans:2016opp,Bhattacharyya:2017hdc}

\begin{equation} 
\begin{aligned}
\left.\frac{d N}{d y}\right|_{y=0} &=\frac{g V}{(2 \pi)^{2}} \int_{0}^{\infty} p_{T} d p_{T} m_{T}\left[1+(q-1) \frac{m_{T}}{T}\right]^{-\frac{q}{q-1}} \\
&=\frac{g V T}{(2 \pi)^{2}}\left[\frac{(2-q) m_{0}^{2}+2 m_{0} T+2 T^{2}}{(2-q)(3-2 q)}\right]\left[1+(q-1) \frac{m_{0}}{T}\right]^{-\frac{1}{q-1}}, \label{eqb:3}
\end{aligned}
\end{equation}

where $m_{0}$ stands for the mass of the used particle.

From Eq. (\ref{eqb:3}), the volume of the system can be written in terms of the multiplicity per rapidity $d N / d y$ and the Tsallis parameters $q$ and $T$ as

\begin{equation} 
V=\left.\frac{d N}{d y}\right|_{y=0} \frac{(2 \pi)^{2}}{g T}\left[\frac{(2-q)(3-2 q)}{(2-q) m_{0}^{2}+2 m_{0} T+2 T^{2}}\right]\left[1+(q-1) \frac{m_{0}}{T}\right]^{\frac{1}{q-1}}, \label{eqb:4}
\end{equation}

Substituting from Eq. (\ref{eqb:4}) into Eq. (\ref{eqb:3}), one obtains the transverse momentum spectra 

\begin{equation} 
\begin{aligned}
\left.\frac{1}{p_{T}} \frac{d^{2} N}{d p_{T} d y}\right|_{y=0}=&\left.\frac{d N}{d y}\right|_{y=0} \frac{m_{T}}{T} \frac{(2-q)(3-2 q)}{(2-q) m_{0}^{2}+2 m_{0} T+2 T^{2}}\left[1+(q-1) \frac{m_{0}}{T}\right]^{\frac{1}{q-1}} \\
&\left[1+(q-1) \frac{m_{T}}{T}\right]^{-\frac{q}{q-1}}.  \label{eqb:5}
\end{aligned}
\end{equation}
where $d N/d y$, $\mathtt{T}$, and $\mathtt{q}$ are the fitting parameters.

\section{The transverse momentum distribution based on ANN model}
\label{sec:(append:neural)}
\renewcommand{\theequation}{D.\arabic{equation}}
\setcounter{equation}{0}
the transeverse momentum distribution $\frac{1}{N_{evt} }\frac{d^2 N}{ d y d p_T}$, can be estimated from ANN model as:

\begin{equation}
\begin{aligned}
\frac{1}{N_{evt} }\frac{d^2 N}{ d y d p_T} &= \text{purelin}[net.LW\left\{5,4\right\}f(net.LW\left\{4,3\right\}f(net.LW\left\{3,2\right\}f(net.LW\left\{2,1\right\} \\
	& f(net.IW\left\{1,1\right\}R+net.b\left\{1\right\})+ net.b\left\{2\right\})+ net.b\left\{3\right\})\\
&	+ net.b\left\{4\right\})+ net.b\left\{5\right\}]. 
\label{equ:seventeenn}
\end{aligned} 	
\end{equation}

Where $R$ is the inputs ( $\sqrt{S} $, $p_{T}$ and centrality),\newline
$f$ is hidden layer transfer function (logsig or poslin),\newline
$IW$ and $LW$ are the linked weights as follow: \newline
$net.IW \left\{1, 1\right\}$ is linked weights between the input layer and first hidden layer,\newline
$net.LW \left\{2, 1\right\}$ is linked weights between first and second hidden layer,\newline
$net.LW \left\{3, 2\right\}$ is linked weights between the second and third hidden layer,\newline
$net.LW \left\{4, 3\right\}$ is linked weights between the third and fourth hidden layer,\newline
$net.LW \left\{5, 4\right\}$ is linked weights between the fourth and output layer,\newline
and $b$ is the bias and considers as follow:\newline
$net.b\left\{1\right\}$ is the bias of the first hidden layer,\newline
$net.b\left\{2\right\}$ is the bias of the second hidden layer,\newline
$net.b\left\{3\right\}$ is the bias of the third hidden layer,\newline
$net.b\left\{4\right\}$ is the bias of the fourth hidden layer,and \newline
$net.b\left\{5\right\}$ is the bias of the output layer.

\end{appendices}

\section{References}

\bibliographystyle{aip}

\bibliography{Quantum_Correction_with_Ratios}

\end{document}